\documentclass[twocolumn,showpacs,reprint,amsmath,amssymb,eqsecnum,pre,showpacs,aps]{revtex4-1}

\usepackage{graphicx}% Include figure files
\usepackage{dcolumn}% Align table columns on decimal point
\usepackage{bm}% bold math

    \usepackage[usenames]{color}

\newcommand{\beq}{\begin{equation}}
\newcommand{\eeq}{\end{equation}}

\newcommand\beqa{\begin{eqnarray}}
\newcommand\eeqa{\end{eqnarray}}
\newcommand{\nn}{\nonumber\\}
\newcommand{\dd}{\text{d}}
\newcommand{\al}{\alpha}

\newcommand{\ot}{\widetilde{t}}
\newcommand{\on}{\widetilde{n}}
\newcommand{\oT}{\widetilde{T}}
\newcommand{\ov}{\widetilde{\mathbf{v}}}
\newcommand{\of}{\widetilde{f}}
\newcommand{\onu}{\widetilde{\nu}}
\newcommand{\ozeta}{\widetilde{\zeta}}
\newcommand{\oP}{\widetilde{P}}

\newcommand{\wt}{\widehat{t}}
\newcommand{\wv}{\widehat{\mathbf{v}}}
\newcommand{\wf}{\widehat{f}}
\newcommand{\wa}{\widehat{a}}
\newcommand{\wT}{\widehat{T}}

\newcommand{\ed}{\bibliographystyle{apsrev}\bibliography{D:/bib_files/Granular}\end{document}}

\newcommand{\h}{\ell}
\newcommand{\lb}{\theta}
\newcommand{\lc}{\vartheta}

\begin{document}
\title{Unsteady non-Newtonian hydrodynamics in  granular gases}

\author{Antonio Astillero}
\email{aavivas@unex.es}
\homepage{\url{http://www.unex.es/eweb/fisteor/antonio_astillero/}}
\affiliation{Departamento de Tecnolog\'ia de Computadores y Comunicaciones,  Universidad de Extremadura, E--06800 M\'erida, Spain}
\author{Andr\'es Santos}
\email{andres@unex.es}
\homepage{\url{http://www.unex.es/eweb/fisteor/andres}}
\affiliation{Departamento de F\'{\i}sica, Universidad de Extremadura,
E-06071 Badajoz, Spain}

\date{\today}
\begin{abstract}
The temporal evolution of a dilute granular gas, both in a compressible flow (uniform longitudinal flow) and in an incompressible flow (uniform shear flow), is investigated by means of the direct simulation Monte Carlo method to solve the Boltzmann equation. Emphasis is laid on the identification of a first ``kinetic'' stage (where the physical properties are strongly dependent on the initial state) subsequently followed by an unsteady ``hydrodynamic'' stage (where the momentum fluxes are well-defined non-Newtonian functions of the rate of strain). The simulation data are seen to support this two-stage scenario. Furthermore, the rheological functions obtained from simulation are  well described by an approximate analytical solution of a model kinetic equation.

\end{abstract}


%02.50.-r, 05.20.Jj, 51.30.+i} % PACS, the Physics and Astronomy
                                    % Classification Scheme.
      %02.50.-r Probability theory, stochastic processes and statistics
      %05.20.Jj Statistical mechanics of classical fluids
      %34.20.Cf Interatomic potentials and forces
      %51.30.+i Thermodynamic properties, equations of state
      %52.25.Jm Ionization of plasmas
%\keywords{Fluids, additive mixtures, hard particles, high Euclidean dimensions}
%Use showkeys class option if keyword display desired

\maketitle

%%%%%%%%%%%%%%%%%%%%%%%%%%%%%%%%%%%%%%%%%%%%%%%%%%%%%
\section{Introduction and statement of the problem \label{sec1}}

Granular gases are dilute systems made of inelastic particles that can be maintained in a fluidized state by the application of external drivings to
compensate for the dissipation of kinetic energy due to collisions. These systems are always out of equilibrium and exhibit a wealth of
intriguing complex phenomena \cite{C90,JNB96a,JNB96b,K99,OK00,K00,D00,PL01,D02,HM02,G03,BP04}.
They are important from an applied point of view but also at the level of fundamental physics. As Kadanoff stated in his review paper \cite{K99}, ``one might even say that the study of
granular materials gives one a chance to reinvent statistical mechanics in a new context.''

One of the most controversial issues in granular fluids refers to the validity of a hydrodynamic description \cite{K99,TG98}. In conventional fluids, the densities of
the conserved quantities (mass, momentum, and energy) satisfy formally exact balance (or continuity) equations involving the divergence of the associated fluxes.
In the case of granular fluids, however, energy is dissipated on collisions and this  gives rise to a sink term in the energy balance
equation.  As a consequence, except perhaps in quasielastic
situations, the role of the energy density (or, equivalently, of the granular temperature) as a hydrodynamic
variable is not  evident. Both for conventional and granular fluids, the mass, momentum, and energy balance equations do not form a closed set due to the appearance of the momentum and energy fluxes (plus the energy sink in the granular case). On the other hand, by assuming ``hydrodynamic'' conditions, the balance equations are closed by the addition of approximate constitutive equations relating the momentum and energy fluxes (again plus the energy sink in the granular case) to the mass, momentum, and energy fields.

The simplest constitutive equations consist of replacing the fluxes by their local equilibrium forms, thus neglecting the influence of the hydrodynamic gradients. This gives rise to the Euler hydrodynamic equations, which fail to account for irreversible effects, even in the case of conventional fluids. This is corrected by the Navier--Stokes (NS) constitutive equations, where the fluxes
are assumed to be linear in the hydrodynamic gradients. On the other hand, if the gradients are not weak enough (i.e., if the Knudsen number is not small enough), the NS equations are insufficient and, thus, nonlinear (i.e., non-Newtonian) constitutive equations are needed in a hydrodynamic description \cite{CC70,GS03}.

\begin{figure}
\includegraphics[width=.9 \columnwidth]{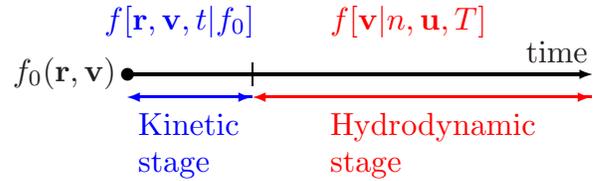}
\caption{(Color online) Schematic description of the two-stage evolution of the velocity distribution function in a conventional gas.}
\label{fig1}
\end{figure}

In conventional fluids, applicability of a hydrodynamic description (Euler, NS, or non-Newtonian) requires two  basic conditions, one spatial and another one temporal. On the one hand, one must focus on the \emph{bulk} region of the system, i.e., outside the \emph{boundary layers}, whose width is on the order of the mean free path. On the other hand, one must let the system \emph{age} beyond the \emph{initial layer}, whose duration is on the order of the mean free time. Let us consider the latter condition in more detail.
In a conventional gas,  the typical evolution scenario  starting from an arbitrary initial state represented by an arbitrary initial  velocity distribution $f_0(\mathbf{r},\mathbf{v})$ proceeds along two successive stages \cite{DvB77}. First,
during the so-called \emph{kinetic} stage, the velocity distribution $f(\mathbf{r},\mathbf{v},t)$, which depends functionally on $f_0$, experiences a fast relaxation (lasting a few collision times)  toward a  ``normal''  form $f[\mathbf{v}|n,\mathbf{u},T]$, where all the spatial and temporal dependence occurs through  a
\emph{functional} dependence on the hydrodynamic fields (number density $n$, flow velocity $\mathbf{u}$, and temperature $T$). Next, during the \emph{hydrodynamic} stage, a slower evolution of the hydrodynamic fields takes place until either  equilibrium or an externally imposed nonequilibrium
steady state is eventually reached. While the first stage is very sensitive to the initial preparation of the system, the details of the initial
state are practically ``forgotten''   in the hydrodynamic regime. Figure \ref{fig1} depicts a schematic summary of this two-stage evolution in a conventional gas.

The absence of energy conservation in  granular fluids sheds some reasonable doubts on the applicability of the above scenario beyond the quasielastic  regime. While the usefulness of a \emph{non-Newtonian} hydrodynamic description in \emph{steady states} has been validated by computer simulations \cite{BRM97,TTMGSD01,VSG10,VGS11}, it is not obvious that a hydrodynamic treatment holds as well during the \emph{transient} regime toward the steady state.

\begin{figure}
\includegraphics[width=.7 \columnwidth]{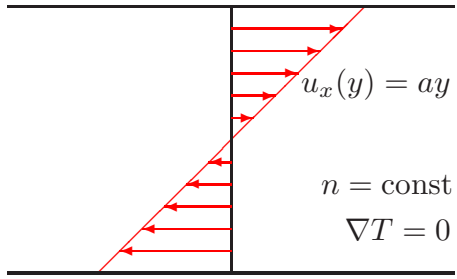}
\caption{(Color online) Sketch of the USF.}
\label{fig2}
\end{figure}
\begin{figure}
\includegraphics[width=.9 \columnwidth]{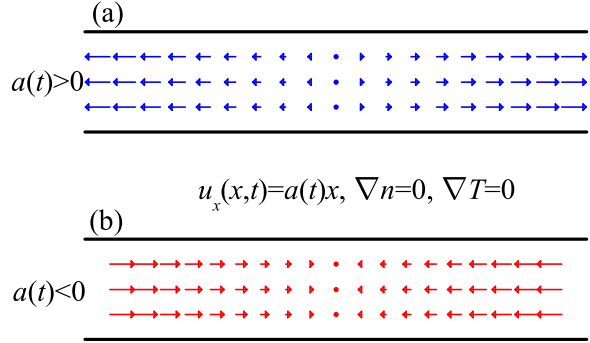}
\caption{(Color online) Sketch of the ULF for (a) $a(t)>0$ and (b) $a(t)<0$.}
\label{fig3}
\end{figure}

In order to address the problem described in the preceding paragraph, it seems convenient to focus on certain prototypical classes of flows. Let us assume a ($d$-dimensional) granular gas with uniform density $n(t)$, uniform temperature $T(t)$, and a flow velocity along a given axis (say $x$) with a linear spatial variation with respect to a certain Cartesian coordinate $\h$, i.e.,
\beq
\nabla_j u_i(\mathbf{r},t)=a(t)\delta_{ix}\delta_{j\h}.
\label{1}
\eeq
Here $a(t)$ is a uniform rate of strain. Two distinct possibilities arise: either $\h\neq x$ (say $\h=y$) or $\h=x$. The first case defines an
incompressible flow ($\nabla\cdot \mathbf{u}=0$) commonly known as simple or uniform shear flow (USF)
\cite{TM80,LSJC84,CG86,WB86,JR88,C89,C90,S92,HS92,SK94,LB94,GT96,SGN96,C97,BRM97,CR98,MGSB99,F00,K00a,%
K00b,K01,C00,C01,C01b,AH01,CH02,BGS02,G02,MG02,MG03,GM03,G03a,GS03,AL03a,AL03b,G03,SGD04,L04,AS05,MGAS06,SG07,AS07,S08,S08a}, the associated rate of strain $a$ being the \emph{shear rate}. The second case is an example of compressible flow ($\nabla\cdot \mathbf{u}=a\neq 0$) that will be referred to as uniform longitudinal flow (ULF) \cite{GK96,KDN97,UP98,KDN98,UG99,S00a,S08a,S09}; the corresponding rate of strain  in this case will be called \emph{longitudinal rate}. These two states are particular cases of a more general class of
homo-energetic affine flows characterized by $\partial^2
u_i/\partial x_j\partial x_k=0$ \cite{TM80}. The USF and ULF flows are sketched in Figs.\ \ref{fig2} and \ref{fig3}, respectively.

Assuming  that the velocity distribution function $f(\mathbf{r},\mathbf{v},t)$ depends on the spatial variable $\h$ only, the Boltzmann equation reads
\beq
\frac{\partial f}{\partial t}+v_\h\frac{\partial f}{\partial \h}=J[f,f],
\label{2}
\eeq
where $J[f,f]$ is the (inelastic) Boltzmann collision operator, whose explicit form can be found, for instance, in Refs.\ \cite{GS95,DBS97,NE01}.
 {Multiplying both sides of Eq.\ \eqref{2} by $\{1,\mathbf{v},v^2\}$, and integrating over velocity, we get the balance equations for mass, momentum, and energy densities},
\begin{equation}
 {D_tn=-n\frac{\partial u_\h}{\partial \h} ,}
\label{nbal}
\end{equation}
\begin{equation}
 {D_t{u_i}=-\frac{1}{mn}\frac{\partial P_{\h i}}{\partial\h} ,}
\label{Pbal}
\end{equation}
\begin{equation}
 {D_tT+\zeta T=-\frac{2}{dn}\left(P_{\h i}\frac{\partial u_i}{\partial \h}+\frac{\partial q_\h}{\partial \h}\right)  .}
\label{Tbal}
\end{equation}
 {In Eqs.\ \eqref{nbal}--\eqref{Tbal}, $D_t\equiv \partial_t+{u}_\h\partial_\h$ is the material derivative, and the number density $n$, flow velocity $\mathbf{u}$, temperature $T$,  pressure tensor $P_{ij}$, heat flux vector $\mathbf{q}$, and cooling rate $\zeta$ are defined by}
\beq
n(\h,t)=\int \dd\mathbf{v}\, f(\h,\mathbf{v},t),
\label{n}
\eeq
\beq
n({\h},t)\mathbf{u}({\h},t)=\int \dd{\mathbf{v}}\,{\mathbf{v}}{f}({\h},{\mathbf{v}},{t}),
\label{u}
\eeq
\beq
n({\h},t)T({\h},t)=p({\h},t)=\frac{1}{d}\text{tr}\mathsf{P}({\h},t),
\label{T}
\eeq
\beq
P_{ij}({\h},t)={m}\int \dd{\mathbf{v}}\,[{v_i}-{u_i}({\h},t)][{v_j}-{u_j}({\h},t)]{f}({\h},{\mathbf{v}},{t}),
\label{PPij}
\eeq
\beq
 {\boldsymbol{q}({\h},t)=\frac{m}{2}\int \dd\mathbf{v}\,[\mathbf{v}-\mathbf{u}({\h},t)]^2[\mathbf{v}-\mathbf{u}({\h},t)]f({\h},\mathbf{v},t),}
\label{q}
\eeq
\beq
 {n({\h},t)T({\h},t)\zeta({\h},t)=-\frac{m}{d}\int \dd\mathbf{v}\,v^2 J[f,f].}
\label{zeta}
\eeq
 {The first equality of Eq.\ \eqref{T} defines the hydrostatic pressure $p$, which is just given by the ideal-gas law in the Boltzmann limit.}

As said above, the density $n(t)$ and temperature $T(t)$, and so the hydrostatic pressure $p(t)$, are uniform in the (fully developed) USF and ULF. On physical grounds, it can also be assumed that the whole pressure tensor $P_{ij}(t)$ is uniform as well. Moreover, in the absence of thermal and density gradients, the heat flux can be expected to vanish. Taking all of this into account,  {as well as Eq.\ \eqref{1}}, the balance equations [Eqs.\ {\eqref{nbal}--\eqref{Tbal}}] become
\beq
\dot{n}(t)=-n(t)a(t)\delta_{x\h},
\label{3}
\eeq
\beq
\dot{a}(t)=-a^2(t)\delta_{x\h},
\label{4}
\eeq
\beq
\dot{T}(t)=-\frac{2a(t)}{dn(t)}P_{x\h}(t)-\zeta(t)T(t).
\label{5}
\eeq
In the case of the USF ($\h=y$), Eqs.\ \eqref{3} and \eqref{4} imply that both the density and the shear rate   are constant quantities. As for the temperature, it evolves in time subject to two competing effects: viscous heating [represented by the term $-(2/dn)aP_{xy}>0$] and inelastic cooling [represented by $\zeta T$]. Both effects eventually cancel each other in the steady state.

In the case of the ULF ($\h=x$), the solution to Eqs.\ \eqref{3} and \eqref{4} is
\beq
\frac{n(t)}{a(t)}=\frac{n_0}{a_0},\quad a(t)=\frac{a_0}{1+a_0 t},
\label{6}
\eeq
where $n_0$ is the initial density and $a_0$ is the initial longitudinal rate.
In contrast
to the USF, the sign of $a(t)$ (or, equivalently, the sign of $a_0$) plays a relevant role
and defines two separate situations (see Fig.\ \ref{fig3}). The
case $a>0$ corresponds to a progressively  slower {\em
expansion\/} of the gas from the plane $x=0$ into all of space.  On
the other hand, the case $a<0$ corresponds to a progressively
faster {\em compression\/} of the gas toward the plane $x=0$. The
latter takes place over a {\em finite\/} time period $t=|a_0|^{-1}$.
However, since the collision frequency rapidly increases with time,
the finite period $t=|a_0|^{-1}$ comprises an {\em infinite\/}
number of collisions per particle \cite{S09}.
Given that $P_{xx}>0$, the energy balance equation [see Eq.\ \eqref{5}]  implies that the temperature monotonically decreases with time in the ULF with $a>0$. On the other hand, if $a<0$, we have again a competition between viscous heating and inelastic cooling, so the temperature either increases or decreases (depending on the initial state) until a steady state is eventually reached.
The main characteristic features of the USF and the ULF are summarized in Table \ref{table1}.

\begin{table}
\caption{Main characteristic features of the USF and the ULF.\label{table1}}
\begin{ruledtabular}
\begin{tabular}{cccc}
&USF&ULF ($a>0$)&ULF ($a<0$)\\
\hline
Inelastic cooling&Yes&Yes&Yes\\
Viscous heating&Yes&No&Yes\\
$T(t)\downarrow$ \& $|a^*(t)|\uparrow$&Yes, if&Yes&Yes, if\\
&$\zeta>\frac{2|aP_{xy}|}{dp}$&&$\zeta>\frac{2|a|P_{xx}}{dp}$\\
$T(t)\uparrow$ \& $|a^*(t)|\downarrow$&Yes, if&No&Yes, if\\
&$\zeta<\frac{2|aP_{xy}|}{dp}$&&$\zeta<\frac{2|a|P_{xx}}{dp}$\\
Steady state&Yes&No&Yes\\
\end{tabular}
\end{ruledtabular}
\end{table}

Regardless of whether the rate of strain $a$ is constant (USF) or changes with time (ULF), the relevant parameter is the ratio
\beq
a^*(t)=\frac{a(t)}{\nu(t)}
\label{a*}
\eeq
between $a(t)$ and a characteristic collision frequency $\nu(t)\propto n(t)[T(t)]^{1/2}$. Note that the absolute value of $a^*(t)$ represents the Knudsen number of the problem, i.e., the ratio between the mean free path and the characteristic length associated with the velocity gradient \cite{S08,S08a}. Since $a(t)/n(t)=\text{const}$ both in the USF and the ULF, we have $a^*(t)\propto [T(t)]^{-1/2}$. Consequently, the qualitative behavior of $|a^*(t)|$ is the opposite to that of $T(t)$, as indicated in Table \ref{table1}.

The scenario depicted in Fig.\ \ref{fig1}, if applicable to a granular gas in the USF or in the ULF, means that, after the kinetic stage, the velocity distribution function $f[\mathbf{r},\mathbf{v},t|f_0]$ should adopt a hydrodynamic (or normal) form
\beq
f[\mathbf{r},\mathbf{v},t|f_0]\to n(t)\left[\frac{m}{2T(t)}\right]^{d/2}f^*\left(\mathbf{C}(\mathbf{r},t), a^*(t)\right),
\label{7}
\eeq
where
\beq
\mathbf{C}(\mathbf{r},t)=\frac{\mathbf{v}-\mathbf{u}(\mathbf{r},t)}{\sqrt{2T(t)/m}}
\label{8}
\eeq
is the peculiar velocity scaled with the thermal speed.
For a given value of the
coefficient of restitution,
the scaled velocity distribution
function $f^*(\mathbf{C},a^*)$ must be  independent of the details of the initial state $f_0$ and depend on the applied shear or longitudinal rate $a(t)$ through the reduced scaled
quantity $a^*$ only. In other words, if a hydrodynamic description
is possible, the  form \eqref{7} must ``attract'' the manifold of solutions $f[\mathbf{r},\mathbf{v},t|f_0]$ to the
Boltzmann equation \eqref{2} for sufficiently long times, even before the steady state (if it exists) is reached.
Equation \eqref{7} has its counterpart at the level of the velocity
moments. In particular, the pressure tensor $P_{ij}[t|f_0]$ would
become
\beq
P_{ij}[t|f_0]\to n(t)T(t)P_{ij}^*(a^*(t))
\label{10}
\eeq
with well-defined hydrodynamic functions $P_{ij}^*(a^*)$.

A few years ago we reported a preliminary study \cite{AS07} where the validity of the unsteady hydrodynamic forms \eqref{7} and \eqref{10} for the USF was confirmed by means of the direct simulation Monte Carlo  (DSMC) method to solve the Boltzmann equation and by a simple rheological model. The aim of the present paper is twofold. On the one hand, we want to revisit the USF case by presenting a more extensive and efficient set of simulations, by providing a detailed derivation of the rheological model (which was just written down without derivation in Ref.\ \cite{AS07}), and by including  in the analysis the second viscometric function (which was omitted in Ref.\ \cite{AS07}). On the other hand, we perform a similar analysis (both computational and theoretical) in the case of the ULF. This second study is relevant because, despite the apparent similarity between the USF and the ULF, the latter differs from the former in that it is compressible, the two signs of $a$  physically differ, and a steady state is  possible only for negative $a$.

The remainder of the paper is organized as follows. The formal kinetic theory description for both types of flow is presented in Sec.\ \ref{sec2} within a unified framework. Next, Sec.\ \ref{sec3} offers a more specific treatment based on a simple model kinetic equation. In particular, a fully analytical rheological model is derived. Section \ref{sec4} describes the simulation method employed to solve the Boltzmann equation and the classes of initial conditions considered. The most relevant part of the paper is contained in Sec.\ \ref{sec5}, where the results obtained from the simulations are presented and discussed. It is found that the scenario depicted by Fig.\ \ref{fig1} and Eqs.\ \eqref{7} and \eqref{10} is strongly supported by the simulations. Moreover, the simple analytical rheological model is seen to agree quite well with the simulation results. The paper is closed in Sec.\ \ref{sec6} with a summary and conclusions.

\section{Boltzmann equation for USF and ULF\label{sec2}}

Let us consider a granular gas modeled as a system of smooth inelastic hard spheres (of mass $m$, diameter $\sigma$, and constant coefficient of normal restitution $\al$), subject to the USF or to the ULF sketched in Figs.\ \ref{fig2} and \ref{fig3}, respectively.
In the dilute regime, the velocity distribution function $f(\h,\mathbf{v},t)$ obeys the  Boltzmann equation \eqref{2}. As is well known, the adequate boundary conditions for the USF are  Lees--Edwards's boundary conditions \cite{LE72}, which are not but periodic boundary conditions in the comoving Lagrangian frame \cite{DSBR86}. The appropriate boundary conditions for the ULF are much less obvious. In order to construct them, it is convenient to perform a series of mathematical changes of variables. Also, we will proceed by encompassing the ULF and the USF in a common framework.

\subsection{Changes of variables}

We start by defining scaled time and spatial variables $\widetilde{t}$ and  $\widetilde{\h}$ as
\beq
\widetilde{t}=\lb(t),\quad \widetilde{\h}=\dot{\lb}(t)\h,
\label{11}
\eeq
where
\beq
\lb(t)=\begin{cases}
t,&\text{USF } (\h=y),\\
a_0^{-1}\ln(1+a_0 t),&\text{ULF } (\h=x),
\end{cases}
\label{12}
\eeq
\beq
\dot{\lb}(t)=\begin{cases}
1,&\text{USF } (\h=y),\\
(1+a_0 t)^{-1},&\text{ULF } (\h=x).
\end{cases}
\label{13}
\eeq
Also, a new velocity variable $\widetilde{\mathbf{v}}$ is defined as
\beq
\widetilde{\mathbf{v}}=\mathbf{v}-a(t)\h \mathbf{e}_x,
\label{14}
\eeq
where $\mathbf{e}_x$ is the unit vector in the $x$ direction and we have taken into account that $a(t)=a_0\dot{\lb}(t)$ both in the USF and in the ULF.
The velocity distribution function corresponding to the variables $\widetilde{\h}$, $\widetilde{\mathbf{v}}$, and $\widetilde{t}$ is
\beq
\widetilde{f}(\widetilde{\h},\widetilde{\mathbf{v}},\widetilde{t})=\frac{1}{\dot{\lb}(t)}f(\h,\mathbf{v},t).
\label{15n}
\eeq
Consequently,
\beq
\frac{1}{\dot{\lb}^2}\frac{\partial f}{\partial t}=\frac{\partial \widetilde{f}}{\partial \widetilde{t}}-a_0\delta_{x\h}\left[\widetilde{f}+\dot{\lb}{\h}\left(
\frac{\partial \widetilde{f}}{\partial \widetilde{\h}}-a_0\frac{\partial \widetilde{f}}{\partial \widetilde{v}_x}\right)\right],
\label{17}
\eeq
where we have taken into account that
\beq
\ddot{\lb}(t)=-a_0[\dot{\lb}(t)]^2\delta_{x\h}.
\label{16}
\eeq
Similarly,
\beq
\frac{1}{\dot{\lb}^2}\frac{\partial f}{\partial \h}=\frac{\partial \widetilde{f}}{\partial \widetilde{\h}}-a_0\frac{\partial \widetilde{f}}{\partial \widetilde{v}_x},
\label{18}
\eeq
\beq
\frac{1}{\dot{\lb}^2} J[f,f]=J[\widetilde{f},\widetilde{f}].
\label{19a}
\eeq
Inserting Eqs.\ \eqref{17}, \eqref{18}, and \eqref{19a} into Eq.\ \eqref{2}, and taking into account Eq.\ \eqref{14}, one finally gets
\beq
\frac{\partial \widetilde{f}}{\partial \widetilde{t}}+\widetilde{v}_\h\frac{\partial \widetilde{f}}{\partial \widetilde{\h}}-a_0\frac{\partial}{\partial \widetilde{v}_x}\left(\widetilde{v}_\h \widetilde{f}\right)=J[\widetilde{f},\widetilde{f}].
\label{19}
\eeq

It is important to remark that no assumption has been made. Therefore,
Eqs.\ \eqref{2} and \eqref{19} are mathematically equivalent, so  any solution to Eq.\ \eqref{2} can be mapped onto a solution to Eq.\ \eqref{19} and vice versa. While Eq.\ \eqref{2} describes a gas in the absence of external forces, Eq.\ \eqref{19} describes a gas under the influence of a \emph{nonconservative} external force $\widetilde{\mathbf {F}}=-ma_0 \widetilde{v}_\h\mathbf{e}_x$.
Note that in the case of the ULF with $a_0<0$ the finite time interval $0<t<|a_0|^{-1}$ translates into the infinite scaled time interval $0<\widetilde{t}<\infty$.

The density $\widetilde{n}(\widetilde{\h},\ot)$, flow velocity $\widetilde{\mathbf{u}}(\widetilde{\h},\ot)$, temperature $\widetilde{T}(\widetilde{\h},\ot)$, and pressure tensor $\widetilde{P}_{ij}(\widetilde{\h},\ot)$ associated with the scaled distribution \eqref{15n} are defined analogously to Eqs.\ \eqref{n}--\eqref{PPij}. The  quantities with and without tilde are related by
\beq
\widetilde{n}(\widetilde{\h},\ot)=\frac{n(\h,t)}{\dot{\lb}(t)} , \quad \widetilde{P}_{ij}(\widetilde{\h},\ot)=\frac{P_{ij}(\h,t)}{\dot{\lb}(t)},
\label{20n}
\eeq
\beq
\widetilde{\mathbf{u}}(\widetilde{\h},\ot)=\mathbf{u}(\h,t)-a(t) \h \mathbf{e}_x,\quad \widetilde{T}(\widetilde{\h},\ot)=T(\h,t).
\label{21n}
\eeq

At a microscopic level, we now define the USF ($\h=y$) and the ULF ($\h=x$) as spatially \emph{uniform} solutions to Eq.\ \eqref{19}, i.e.,
\beq
\widetilde{f}(\widetilde{\h},\widetilde{\mathbf{v}},\widetilde{t})=\widetilde{f}(\widetilde{\mathbf{v}},\widetilde{t}).
\label{22n}
\eeq
Thus, conservation of mass and momentum implies that $\widetilde{n}=\text{const}$ and $\widetilde{\mathbf{u}}=\text{const}$.
Without loss of generality we can take $\widetilde{\mathbf{u}}=0$.
It seems quite natural that periodic boundary conditions (at $\widetilde{\h}=\pm\widetilde{L}/2$)  are the appropriate ones to complement the (scaled) Boltzmann equation \eqref{19} in order to ensure the consistency with uniform solutions \eqref{22n}, i.e.,
\beq
\widetilde{f}(-\widetilde{L}/2,\widetilde{\mathbf{v}},\widetilde{t})=\widetilde{f}(\widetilde{L}/2,\widetilde{\mathbf{v}},\widetilde{t}).
\label{23}
\eeq

Assuming uniform solutions of Eq.\ \eqref{19} and going back to the original variables, Eqs.\ \eqref{20n} and \eqref{21n} yield $n(t)=\dot{\lb}(t)\widetilde{n}$,
$\mathbf{u}(\h,t)=a(t)\h \mathbf{e}_x$, $T(t)=\widetilde{T}(\widetilde{t})$, and $P_{ij}(t)=\dot{\lb}(t)\widetilde{P}_{ij}(\widetilde{t})$.
Therefore,  uniform solutions to Eq.\ \eqref{19} map onto USF ($\h=y$) or ULF ($\h=x$) solutions to Eq.\ \eqref{2}.
The periodic boundary conditions \eqref{23} translate into
\beq
f\left(-\frac{\widetilde{L}}{2\dot{\lb}(t)},\mathbf{v},t\right)=
f\left(\frac{\widetilde{L}}{2\dot{\lb}(t)},\mathbf{v}+a_0\widetilde{L}\mathbf{e}_x,t\right).
\label{26}
\eeq
In the case of the USF, these are the well-known Lees--Edwards's boundary conditions \cite{LE72}.

While the forms \eqref{2} and \eqref{19} of the Boltzmann equation are fully equivalent, as are the respective boundary conditions \eqref{26} and \eqref{23}, it is obvious that Eqs.\ \eqref{19} and \eqref{23} are much simpler to implement in computer simulations than Eqs.\ \eqref{2} and \eqref{26}. This is especially important if one restricts oneself to uniform solutions of the form \eqref{22n}. In that case, Eq.\ \eqref{19} becomes
\beq
\frac{\partial \widetilde{f}}{\partial \widetilde{t}}-a_0\frac{\partial}{\partial \widetilde{v}_x}\left(\widetilde{v}_\h \widetilde{f}\right)=J[\widetilde{f},\widetilde{f}].
\label{31}
\eeq
The corresponding energy balance equation is
\beq
\dot{\widetilde{T}}(\widetilde{t})=-\frac{2a_0}{d\widetilde{n}}\widetilde{P}_{x\h}(\widetilde{t})-
\widetilde{\zeta}(\widetilde{t})\widetilde{T}(\widetilde{t}),
\label{32}
\eeq
where
\beq
\widetilde{\zeta}=-\frac{m}{d\widetilde{n}\widetilde{T}}\int\dd\widetilde{\mathbf{v}}\, \widetilde{v}^2 J[\widetilde{f},\widetilde{f}].
\label{33}
\eeq
Note that
\beq
\zeta(t)=\dot{\lb}(t)\widetilde{\zeta}(\widetilde{t})
\label{25b}
\eeq
and, thus, Eqs.\ \eqref{5} and \eqref{32} are equivalent.
Although $T=\oT$, in Eq.\ \eqref{32} we keep the notation $\oT$ to emphasize that here the temperature is seen as a function of the scaled time $\ot$.

In cooling situations, i.e., in the USF if $\zeta>2|aP_{xy}|/dp$, in the ULF with $a_0<0$ if $\zeta>2|a|P_{xx}/dp$, or in the ULF with $a_0>0$, the temperature can reach values much smaller than the initial one,  {which} can cause technical difficulties (low signal-to-noise ratio) in the simulations. This is especially important in the ULF with $a_0>0$ since no steady state exists and the temperature keeps decreasing without any lower bound. To manage this problem, it is convenient to introduce a velocity rescaling (or thermostat). {}From a mathematical point of view, let us perform the additional change of variables
\beq
\widehat{t}=\lc(\widetilde{t}),\quad \wv=\frac{\ov}{\dot{\lc}(\ot)},
\label{34}
\eeq
\beq
\wf(\wv,\wt)=\left[\dot{\lc}(\ot)\right]^d \of(\ov,\ot),
\label{35}
\eeq
where so far $\lc(\widetilde{t})$ is an arbitrary (positive definite) protocol function. The following identities are straightforward,
\beq
\left[\dot{\lc}(\ot)\right]^{d-1}\frac{\partial\of}{\partial\ot}=\frac{\partial\wf}{\partial\wt}
-\frac{\ddot{\lc}(\ot)}{\left[\dot{\lc}(\ot)\right]^2}\frac{\partial}{\partial\wv}\left(\wv\wf\right),
\label{36}
\eeq
\beq
\left[\dot{\lc}(\ot)\right]^{d-1}J[\of,\of]=J[\wf,\wf].
\label{37}
\eeq
Therefore, Eq.\ \eqref{31} becomes
\beq
\frac{\partial \wf}{\partial \wt}-\wa(\wt)\frac{\partial}{\partial \widehat{v}_x}\left(\widehat{v}_\h \wf\right)-\mu(\wt)\frac{\partial}{\partial\wv}\left(\wv\wf\right)=J[\wf,\wf],
\label{38}
\eeq
where
\beq
\wa(\wt)\equiv \frac{a_0}{\dot{\lc}(\ot)},\quad \mu(\wt)\equiv \frac{\ddot{\lc}(\ot)}{\left[\dot{\lc}(\ot)\right]^2}.
\label{mu}
\eeq
The Boltzmann equation [Eq.\ \eqref{38}] represents the action of a {nonconservative} external force $\widehat{\mathbf {F}}(\wv,\wt)=-m\wa(\wt)\widehat{v}_\h\mathbf{e}_x-m\mu(\wt)\wv$.
The relationship between the granular temperatures defined from $\wf$ and $\of$, respectively, is $\widehat{T}(\wt)=\widetilde{T}(\ot)/\left[\dot{\lc}(\ot)\right]^2$. Thus, the \emph{thermostat} choice $\dot{\lc}(\ot)\propto \left[\widetilde{T}(\ot)\right]^{1/2}$ keeps the rescaled temperature $\widehat{T}(\wt)$ constant. While, at a theoretical level, Eq.\ \eqref{31} is simpler and more transparent than Eq.\ \eqref{38}, the latter is more useful from a computational point of view in cooling situations.

\subsection{Rheological functions}

In order to characterize the non-Newtonian properties of the unsteady USF and ULF, it is convenient to introduce  generalized  transport coefficients.

As is well known, the NS shear viscosity $\eta_\text{NS}$ is defined through the linear constitutive equation
\beq
P_{ij}=p\delta_{ij}-\eta_\text{NS} \left(\nabla_i u_j+\nabla_j u_i-\frac{2}{d}\nabla\cdot \mathbf{u}\delta_{ij}\right),
\label{etaNS}
\eeq
 {where we have taken into account that the NS bulk viscosity is zero in the low-density limit \cite{CC70,BDKS98}. This is especially relevant in compressible flows like the ULF.
Making use of} Eq.\ \eqref{1}, we define (dimensionless) non-Newtonian viscosities $\eta^*(a^*)$ for the USF and the ULF by the relation
\beq
P_{ij}^*(a^*)=\delta_{ij}-\eta^*(a^*)a^*\left(\delta_{ix}\delta_{j\h}+\delta_{jx}\delta_{i\h}-\frac{2}{d}\delta_{x\h}\delta_{ij}\right).
\label{Pij}
\eeq
More specifically, setting $ij=x\h$, Eq.\ \eqref{Pij} yields
\beq
\eta^*(a^*)=\frac{\delta_{x\h}-P_{x\h}^*(a^*)}{a^*\left(1+\frac{d-2}{d}\delta_{x\h}\right)}.
\label{eta}
\eeq
The rheological function $\eta^*(a^*)$  differs in the USF from that in the ULF.  In the latter flow it is related to the \emph{normal} stress $P_{xx}^*$. In that case, by symmetry, one has $P_{xx}^*+(d-1)P_{yy}^*=d$, so 
\beq
P_{xx}^*-P_{yy}^*=-2\eta^*(a^*)a^*.
\label{Pxx-Pyy}
\eeq
Since $0<P_{xx}^*<d$, the viscosity $\eta^*(a^*)$ in the ULF must be positive definite but upper bounded:
\beq
\eta^*(a^*)<\begin{cases}
\frac{d}{2(d-1)a^*},&a^*>0,\\
\frac{d}{2|a^*|},&a^*<0.
\end{cases}
\eeq

In contrast,  in the USF  the viscosity function is related to the \emph{shear} stress $P_{xy}^*$,
\beq
P_{xy}^*=-\eta^*(a^*)a^*.
\label{etaUSF}
\eeq
In this state the normal stress differences  are characterized by the viscometric functions
\beq
\Psi_1^*(a^*)=\frac{P_{yy}^*(a^*)-P_{xx}^*(a^*)}{{a^*}^2},
\label{43}
\eeq
\beq
\Psi_2^*(a^*)=\frac{P_{zz}^*(a^*)-P_{yy}^*(a^*)}{{a^*}^2}.
\label{43bis}
\eeq

\section{Kinetic model and nonlinear hydrodynamics\label{sec3}}

\subsection{Kinetic model}
In order to progress on the theoretical understanding of the USF and the ULF, it is convenient to adopt an extension of the  Bhatnagar--Gross--Krook (BGK) kinetic model \cite{C88}, in which the (inelastic) Boltzmann collision operator $J[f,f]$ is replaced by a simpler form \cite{BDS99,SA05}:
\beq
J[f,f]\to-\beta(\alpha)\nu(f-f_\text{hcs})+\frac{\zeta}{2}\frac{\partial}{\partial \mathbf{v}}\cdot[(\mathbf{v}-\mathbf{u})f].
\label{n1}
\eeq
Here $f_\text{hcs}$ is the local version of the homogeneous cooling state distribution {\cite{BP04}} and
\beq
\nu=
\frac{8\pi^{(d-1)/2}}{(d+2)\Gamma(d/2)}n\sigma^{d-1}\sqrt{\frac{T}{m}}
\label{0.1}
\eeq
is a convenient choice for the effective collision frequency.
The factor $\beta(\alpha)$ can be freely chosen to optimize agreement with the original Boltzmann equation. Although it is not necessary to fix it in the remainder of this section, we will take
\beq
\beta(\alpha)=\frac{1}{2}(1+\alpha)
\eeq
at the end \cite{SA05}. The cooling rate is defined by Eq.\ \eqref{zeta} but here we will take the expression obtained from the Maxwellian approximation, namely
\beq
\zeta=\frac{d+2}{4d}(1-\alpha^2)\nu.
\label{0.1a}
\eeq
 {This is sufficiently accurate from a practical point of view \cite{AS05}, especially at the level of the simple kinetic model \eqref{n1}.}

Using the replacement \eqref{n1}, Eq.\ \eqref{31} becomes
\beq
\frac{\partial \widetilde{f}}{\partial \widetilde{t}}-a_0\frac{\partial}{\partial \widetilde{v}_x}\left(\widetilde{v}_\h \widetilde{f}\right)=-\beta\onu(\of-\of_\text{hcs})+\frac{\ozeta}{2}\frac{\partial}{\partial \ov}\cdot(\ov\of),
\label{3.1}
\eeq
where $\onu$ and $\ozeta$ are given by Eqs.\ \eqref{0.1} and \eqref{0.1a}, respectively, except for the change $n\to \widetilde{n}$  (recall that $T=\widetilde{T}$).
Taking second-order velocity moments on both sides of Eq.\ \eqref{3.1} one gets
\beq
\dot{\oP}_{ij}=-a_0\left(\oP_{j\h}\delta_{ix}+\oP_{i\h}\delta_{jx}\right)-{\ozeta}\,{\oP_{ij}}-\beta\onu\left(\oP_{ij}-\on \oT\delta_{ij}\right).
\label{3.2}
\eeq
{}From the trace of both sides of Eq.\ \eqref{3.2} we recover the exact energy balance equation \eqref{32}. The advantage of the BGK-like model kinetic equation \eqref{3.1} is that it allows one to complement Eq.\ \eqref{32} with a closed set of equations for the elements of the pressure tensor \cite{BRM97,S09}. It is interesting to note that Eq.\ \eqref{3.2}, with  $\beta=(1+\al)[d+1+(d-1)\al]/4d$, can also be derived from the original Boltzmann equation in the Grad approximation \cite{SGD04,VSG10,VGS11}.

As discussed in Sec.\ \ref{sec1} [cf.\ Eq.\ \eqref{10}], the relevant quantity is the \emph{reduced} pressure tensor defined as
\beq
P_{ij}^*(\ot)=\frac{P_{ij}(t)}{n(t)T(t)}=\frac{\oP_{ij}(\ot)}{\on\oT(\ot)}
\label{3.6}
\eeq
Combining Eqs.\ \eqref{32} and \eqref{3.2} we obtain
\beq
\frac{\dot{P}_{ij}^*}{\onu}=-a^*\left(P_{j\h}^*\delta_{ix}+P_{i\h}^*\delta_{jx}\right)+\frac{2a^*}{d}
P_{ij}^*P_{x\h}^*-\beta\left(P_{ij}^*-\delta_{ij}\right),
\label{3.2bis}
\eeq
where, according to Eq.\ \eqref{a*},
\beq
a^*(\ot)=\frac{a(t)}{\nu(t)}=\frac{a_0}{\onu(\ot)}
\label{3.5}
\eeq
is the reduced shear ($\h=y$) or longitudinal ($\h=x$) rate.
Taking $ij=x\h$ and $ij=\h\h$, Eq.\ \eqref{3.2bis} yields
\beq
\frac{\dot{P}_{x\h}^*}{\onu}=-a^*\left(P_{\h\h}^*+P_{x\h}^*\delta_{x\h}\right)+\frac{2a^*}{d}\left(P_{x\h}^*\right)^2-
\beta\left(P_{x\h}^*-\delta_{x\h}\right),
\label{3.3}
\eeq
\beq
\frac{\dot{P}_{\h\h}^*}{\onu}=-2a^*P_{\h\h}^*\delta_{x\h}+\frac{2a^*}{d}P_{\h\h}^*P_{x\h}^*-\beta\left(P_{\h\h}^*-1\right).
\label{3.4}
\eeq
Note that Eq.\ \eqref{3.4} is identical to Eq.\ \eqref{3.3} in the ULF ($\h=x$).
Equations \eqref{3.3} and \eqref{3.4} must be complemented with the evolution equation for $a^*$. We recall that $\onu\propto \oT^q$ and  $a^*\propto \oT^{-q}$ with $q=\frac{1}{2}$. Thus, using Eq.\ \eqref{32} one simply obtains
\beq
\frac{\dot{a}^*}{\onu}=qa^*\left(\frac{2a^*}{d}P_{x\h}^*+\zeta^*\right),
\label{3.7}
\eeq
where, according to Eq.\ \eqref{0.1a},
\beq
\zeta^*=\frac{d+2}{4d}(1-\alpha^2).
\label{3.8}
\eeq
Here we will temporarily view $q$ as a  free parameter, so  the solutions to  Eqs.\ \eqref{3.3}--\eqref{3.7}  depend parametrically on $q$. The exponent $q$  is directly related to the wide class of dissipative gases introduced by Ernst \emph{et al.} \cite{ETB06a,ETB06b,TBE07}.

Equations \eqref{3.3}--\eqref{3.7} constitute a closed set of nonlinear equations for $\{P_{x\h}^*(\ot),P_{\h\h}^*(\ot),a^*(\ot)\}$ that can be numerically solved subject to a given initial condition
\beq
\of_0(\ov)\Rightarrow\{P_{x\h,0}^*,P_{\h\h,0}^*,a^*_0\}.
\label{ini}
\eeq

\subsection{Steady state}
Setting $\dot{P}_{x\h}^*=0$, $\dot{P}_{\h\h}^*=0$, and $\dot{a}^*=0$ in Eqs.\ \eqref{3.3}--\eqref{3.7}, they become a set of three (USF) or two (ULF) independent algebraic equations whose solution provides the steady-state values. In the case of the USF ($\h=y$) the solution is
\beq
|a^*_s|=\sqrt{\frac{d\zeta^*}{2\beta}}(\beta+\zeta^*),
\label{3.9}
\eeq
\beq
P_{xy,s}^*=-\sqrt{\frac{d\beta\zeta^*}{2}}\frac{\text{sgn}(a^*_s)}{\beta+\zeta^*},\quad
P_{yy,s}^*=\frac{\beta}{\beta+\zeta^*}.
\label{3.10}
\eeq
In contrast, the solution for the ULF ($\h=x$) is
\beq
a^*_s=-\frac{d\zeta^*}{2}\frac{\beta+\zeta^*}{\beta+d\zeta^*},
\label{3.11a}
\eeq
\beq
P_{xx,s}^*=\frac{\beta+d\zeta^*}{\beta+\zeta^*}.
\label{3.11b}
\eeq
Note that the steady-state values are independent of $q$. A linear stability analysis in the case of the USF \cite{SGD04} shows that the steady state, Eqs.\ \eqref{3.9} and \eqref{3.10}, is indeed a stable solution of Eqs.\ \eqref{3.3}--\eqref{3.7}. The proof can be easily extended to the ULF.

\subsection{Unsteady hydrodynamic solution}
In the USF ($\h=y$), Eq.\ \eqref{3.2bis}  implies that $(\dot{P}_{zz}^*-\dot{P}_{yy}^*)/\widetilde{\nu}=(2a^*P_{xy}^*/d-\beta)({P}_{zz}^*-{P}_{yy}^*)$. Since, on physical grounds, $a^*P_{xy}^*<0$, we conclude that ${P}_{zz}^*-{P}_{yy}^*=0$ in the hydrodynamic regime. Therefore, according to the kinetic model description, the second viscometric function  identically vanishes, i.e.,
\beq
\Psi_2^*(a^*)=0.
\label{Psi2_0}
\eeq
Next, by symmetry,
$P_{xx}^*+P_{yy}^*+(d-2)P_{zz}^*=d$. This mathematical identity, combined with ${P}_{zz}^*={P}_{yy}^*$, allows one to rewrite Eq.\ \eqref{43} as
\beq
\Psi_1^*=-d\frac{1-{P}_{yy}^*}{{a^*}^2}.
\label{Psi1_0}
\eeq

As sketched in Fig.\ \ref{fig1} and described by Eqs.\ \eqref{7} and \eqref{10}, the hydrodynamic solution requires  the whole time dependence of $P_{ij}^*$ to be captured through a dependence on $a^*$ common to every initial state.
As a first step to obtain such a hydrodynamic solution, let us eliminate time in favor of $a^*$ in Eqs.\ \eqref{3.3} and \eqref{3.4} with the help of Eq.\ \eqref{3.7}, i.e.,
\beqa
q\left(\frac{2a^*}{d}P_{x\h}^*+\zeta^*\right)\frac{\partial{P}_{x\h}^*}{\partial a^*}&=&-P_{\h\h}^*-P_{x\h}^*\delta_{x\h}+\frac{2}{d}\left(P_{x\h}^*\right)^2\nn
&&-\frac{\beta}{a^*}\left(P_{x\h}^*-\delta_{x\h}\right),
\label{3.12}
\eeqa
\beqa
q\left(\frac{2a^*}{d}P_{x\h}^*+\zeta^*\right)\frac{\partial{P}_{\h\h}^*}{\partial a^*}&=&-2P_{\h\h}^*\delta_{x\h}+\frac{2}{d}P_{\h\h}^*P_{x\h}^*\nn
&&-\frac{\beta}{a^*}\left(P_{\h\h}^*-1\right).
\label{3.13}
\eeqa
This set of two nonlinear coupled differential equations  must be solved in general with the initial
conditions stemming from Eq.\ \eqref{ini}, namely
\beq
a^*=a^*_0\Rightarrow \{P_{x\h}^*=P_{x\h,0}^*, P_{\h\h}^*=P_{\h\h,0}^*\}.
\label{3.14}
\eeq
Equations \eqref{3.12} and \eqref{3.13} must be solved in agreement with the physical direction of time. This means that the solutions uncover the region $|a^*_0|<|a^*|<|a_s^*|$ in conditions of cooling  and the region  $|a^*_0|>|a^*|>|a_s^*|$ in conditions of heating (see Table \ref{table1}).
In the case of the ULF with $a^*_0>0$, one has  $a^*_0<a^*<\infty$ due to the absence of steady state.
Equations \eqref{3.12} and \eqref{3.13} describe both the kinetic and hydrodynamic
regimes. In order to isolate the hydrodynamic solution, one must
apply appropriate boundary conditions \cite{SGD04,S09}.

An alternative route to get the hydrodynamic solution consists of expanding $P_{x\h}^*$ and $P_{\h\h}^*$ in powers of $a^*$ (Chapman--Enskog expansion),
\beq
P_{x\h}^*(a^*)=\delta_{x\h}+\sum_{j=1}^\infty c_{x\h}^{(j)}{a^*}^j,\quad P_{\h\h}^*(a^*)=1+\sum_{j=1}^\infty c_{\h\h}^{(j)}{a^*}^j.
\label{3.15}
\eeq
Inserting the expansions in both sides of Eqs.\ \eqref{3.12} and \eqref{3.13} one can get the Chapman--Enskog coefficients $c_{x\h}^{(j)}$ and $c_{\h\h}^{(j)}$ in a recursive way. The first- and second-order coefficients are
\beq
c_{x\h}^{(1)}=-\frac{d+(d-2)\delta_{x\h}}{d(\beta+q\zeta^*)},\quad c_{\h\h}^{(1)}=c_{x\h}^{(1)}\delta_{x\h},
\label{3.16}
\eeq
\beq
c_{x\h}^{(2)}=c_{\h\h}^{(2)}\delta_{x\h},
\label{3.17}
\eeq
\beq
c_{\h\h}^{(2)}=2\frac{2(d-1)(d-2+q)\delta_{x\h}+d(\delta_{x\h}-1)}{d^2(\beta+q\zeta^*)(\beta+2q\zeta^*)}.
\label{3.18}
\eeq
Equation \eqref{3.16} gives NS coefficients, while Eqs.\ \eqref{3.17} and \eqref{3.18} correspond to Burnett coefficients.
{}From Eqs.\ \eqref{3.15} and \eqref{3.16}, Eq.\ \eqref{eta} yields
\beq
\lim_{a^*\to 0}\eta^*(a^*)=\frac{1}{\beta+q\zeta^*}.
\label{eta0NS}
\eeq
Thus, the NS viscosity coincides in the USF and in the ULF, as expected.
Regarding the USF first viscometric function, Eqs.\ \eqref{Psi1_0}, \eqref{3.15}, and \eqref{3.18} gives the Burnett coefficient
\beq
\lim_{a^*\to 0}\Psi_1^*(a^*)=-\frac{2}{(\beta+q\zeta^*)(\beta+2q\zeta^*)}.
\label{Burnett}
\eeq

In general,  all the even (odd) coefficients of $P_{xy}^*$ ($P_{yy}^*$) vanish in the USF. In the ULF, however, all the coefficients of $P_{xx}^*$ are non-zero.
It is interesting to remark that, in contrast to the elastic case ($\zeta^*=0$) \cite{SBD86,S00a}, the  Chapman--Enskog expansions \eqref{3.15} are \emph{convergent} \cite{S08,S08a,S09} if $\zeta^*>0$. On the other hand, the radius of convergence is  finite and coincides with the stationary value $|a_s^*|$.

The series \eqref{3.15} clearly correspond to the hydrodynamic solution since they give $P_{x\h}^*$ and $P_{\h\h}^*$ as unambiguous functions of $a^*$, regardless of the details of the initial conditions \eqref{3.14}. However, the series have two shortcomings. First, since they diverge for $|a^*|>|a_s^*|$, they do not provide $P_{x\h}^*(a^*)$ and $P_{\h\h}^*(a^*)$ in a direct way for that region. Second, even if $|a^*|<|a_s^*|$,  closed expressions for $P_{x\h}^*(a^*)$ and $P_{\h\h}^*(a^*)$ are not possible.

In order to get closed and explicit   (albeit approximate) solutions, we formally take $q$ as a small parameter and
perturb around $q=0$ \cite{KDN97,S00a},
\beq
P_{ij}^*(a^*)=P_{ij}^{*(0)}(a^*)+q P_{ij}^{*(1)}(a^*)+\cdots.
\label{3.19}
\eeq
Setting $q=0$ in Eqs.\ \eqref{3.12} and \eqref{3.13} one gets
\beq
P_{x\h}^{*(0)}=d\left[\delta_{x\h}-\frac{\beta\gamma_\h(a^*)}{a^*}\right],
\label{20}
\eeq
\beq
P_{\h\h}^{*(0)}=\frac{1}{1+2\gamma_\h(a^*)},
\label{21}
\eeq
where $\gamma_\h(a^*)$ is the physical solution of a cubic (USF, $\h=y$) or a quadratic (ULF, $\h=x$) equation:
\beq
\gamma_y(1+2\gamma_y)^2=\frac{{a^*}^2}{\beta^2d},
\label{17USF}
\eeq
\beq
d\left(1-\frac{\beta\gamma_x}{a^*}\right)(1+2\gamma_x)=1.
\label{17ULF}
\eeq
The respective solutions are
\beq
\gamma_y(a^*)=
\frac{2}{3}\sinh^2\left[\frac{1}{6}\cosh^{-1}\left(1+27\frac{{a^*}^2}{\beta^2 d}\right)\right],
\label{15}
\eeq
\beq
\gamma_x(a^*)=\frac{a^*}{2\beta}-\frac{1}{4}+\frac{1}{2}\sqrt{\left(\frac{a^*}{\beta}+\frac{1}{2}\right)^2-\frac{2a^*}{\beta d}}.
\label{15ULF}
\eeq
For small $|a^*|$ one has
\beq
\gamma_y(a^*)=\frac{{a^*}^2}{\beta^2d}+\mathcal{O}({a^*}^4),
\eeq
\beq
\gamma_x(a^*)=\frac{d-1}{d}\frac{{a^*}}{\beta}\left(1+\frac{2}{d}\frac{{a^*}}{\beta}\right)+\mathcal{O}({a^*}^3),
\eeq
so 
\beq
P_{x\h}^{*(0)}(a^*)=\delta_{x\h}-\frac{d+(d-2)\delta_{x\h}}{\beta d}a^*+\mathcal{O}({a^*}^2),
\label{NS0}
\eeq
in agreement with Eq.\ \eqref{3.16}. Furthermore, it is easy to check that in the steady state [cf.\ Eqs.\ \eqref{3.9} and \eqref{3.11a}] one gets
\beq
\gamma_y(a^*_s)=\frac{1}{2}\frac{\zeta^*}{\beta},
\label{22USF}
\eeq
\beq
\gamma_x(a^*_s)=-\frac{d-1}{2}\frac{\zeta^*}{\beta+d\zeta^*},
\label{22ULF}
\eeq
so  Eqs.\ \eqref{20} and \eqref{21} yield Eqs.\ \eqref{3.10} and \eqref{3.11b}, as expected.

Once $P_{ij}^{*(0)}$ are known, Eqs.\ \eqref{3.12} and \eqref{3.13} provide $P_{ij}^{*(1)}$. In the USF case ($\h=y$) the results are
\beq
P_{xy}^{*(1)}=-P_{xy}^{*(0)}h_y(a^*),
\label{31bis}
\eeq
\beq
P_{yy}^{*(1)}=6 P_{yy}^{*(0)}\gamma_y(a^*)H_y(a^*),
\label{32bis}
\eeq
where
\beq
h_y(a^*)\equiv \frac{\left[\zeta^*/\beta-2\gamma_y(a^*)\right]\left[1-6\gamma_y(a^*)\right]}{\left[1+6\gamma_y(a^*)\right]^2},
\label{hUSF}
\eeq
\beq
H_y(a^*)\equiv
\frac{\zeta^*/\beta-2\gamma_y(a^*)}{\left[1+6\gamma_y(a^*)\right]^2},
\label{HUSF}
\eeq
and use has been made of Eq. \eqref{17USF} and the relation
\beq
\frac{\partial \gamma_y(a^*)}{\partial
a^*}=\frac{2a^*}{\beta^2d}\frac{1}{[1+2\gamma_y(a^*)][1+6\gamma_y(a^*)]}.
\label{33b}
\eeq
Analogously, taking into account Eq.\ \eqref{17ULF}, the final expression in the ULF case ($\h=x$) is
\beq
P_{xx}^{*(1)}=
2 P_{xx}^{*(0)}\gamma_x(a^*)h_x(a^*),
\label{32ULF}
\eeq
where
\beq
h_x(a^*)\equiv\frac{2a^*/\beta-2\gamma_x(a^*)+\zeta^*/\beta}{\left[1-2a^*/\beta+4\gamma_x(a^*)\right]^2}.
\label{hULF}
\eeq

Note that in the steady state $P_{ij}^{*(1)}=0$ both in the USF and the ULF. This is consistent with the fact that the steady-state values are independent of $q$. For small $|a^*|$,
\beq
P_{x\h}^{*(1)}(a^*)=\frac{d+(d-2)\delta_{x\h}}{\beta^2 d}\zeta^*a^*+\mathcal{O}({a^*}^2),
\label{NS1}
\eeq
 {which} again agrees with Eq.\ \eqref{3.16}.

In principle, it is possible to proceed further and get the terms of
orders $q^2$, $q^3$,  \ldots, in Eq.\ \eqref{3.19} \cite{S00a}. However, for our purposes it is
sufficient to retain the linear terms only.

\subsection{Rheological model}

The definitions of the rheological functions \eqref{eta},  \eqref{43}, and \eqref{43bis} are independent of any specific model employed to obtain $P_{ij}^*(a^*)$. Here we make use of the kinetic model \eqref{n1} and the expansion \eqref{3.19} truncated to first order in $q$. Next, by means of a Pad\'e approximant we construct  (approximate) explicit expressions for $\eta^*(a^*)$. Let us start with the ULF ($\h=x$), in which case Eqs.\  \eqref{20} and \eqref{32ULF} yield
\beq
\eta^*(a^*)\approx\frac{d}{d-1}\frac{\gamma_x(a^*)}{a^*[1+2\gamma_x(a^*)]}\frac{1}{1+q h_x(a^*)}.
\label{etaULF}
\eeq
Analogously, in the USF case ($\h=y$), Eqs.\ \eqref{20}, \eqref{17USF}, and \eqref{31bis} give
\beq
\eta^*(a^*)\approx\frac{1}{\beta
\left[1+2\gamma_y(a^*)\right]^{2}}\frac{1}{1+qh_y(a^*)}.
\label{etaUSFPade}
\eeq

Let us analyze now the  USF  first viscometric function.
Using Eqs.\ \eqref{21}, \eqref{17USF}, and \eqref{32bis},  Eq.\ \eqref{Psi1_0} gives
\beq
\Psi_1^*(a^*)=-\frac{2}{\beta^{2}
\left[1+2\gamma_y(a^*)\right]^{3}}\left[1-3qH_y(a^*)\right]+\mathcal{O}(q^2).
\eeq
Again, it is convenient to construct a Pad\'e approximant of $\Psi_1^*(a^*)$. Here we take
\beq
\Psi_1^*(a^*)\approx-\frac{2}{\beta^{2}\left[1+2\gamma_y(a^*)\right]^{3}}\frac{1}{[1+q H_y(a^*)][1+2q H_y(a^*)]}.
\label{Psi1}
\eeq
In principle, one should have written $1+3qH_y$ instead of $(1+qH_y)(1+2qH_y)$ in Eq.\ \eqref{Psi1}, but the form chosen  has the advantage of being consistent with the Burnett coefficient \eqref{Burnett} for any $q$.

In summary, our simplified rheological model consists of Eq.\ \eqref{etaULF}, complemented with Eqs.\ \eqref{15ULF}  and \eqref{hULF}, for the ULF and Eqs.\ \eqref{etaUSFPade} and \eqref{Psi1}, complemented with Eqs.\ \eqref{15}, \eqref{hUSF},  and \eqref{HUSF}, for the USF. Since we are interested in hard spheres, we must take $q=\frac{1}{2}$ in those equations.

This approximation has a number of important properties. First,  as said in connection with the Chapman--Enskog expansion \eqref{3.15}, Eqs.\ \eqref{etaULF}, \eqref{etaUSFPade}, and \eqref{Psi1} qualify as a (non-Newtonian) \emph{hydrodynamic} description. Second, in contrast to the full expansions  \eqref{3.15} and \eqref{3.19}, they provide the relevant elements $P_{x\h}^*$ of the pressure tensor as  \emph{explicit} functions of both the shear or longitudinal rate $a^*$ and the coefficient of normal restitution $\alpha$ (through $\beta$ and $\zeta^*$). Third, as seen from Eq.\ \eqref{eta0NS}, Eqs.\ \eqref{etaULF} and \eqref{etaUSFPade}  agree with the \emph{exact} NS coefficients  predicted by the kinetic model \eqref{n1} for arbitrary values of the parameter $q$; this agreement extends to the Burnett-order coefficient \eqref{Burnett}. Next, the correct \emph{steady-state} values [Eqs.\ \eqref{3.9}--\eqref{3.11b}] are included in the Pad\'e approximants \eqref{etaULF}, \eqref{etaUSFPade}, and \eqref{Psi1}. Finally, it can be checked that the correct \emph{asymptotic} forms in the limit $|a^*|\to\infty$ \cite{SGD04,S09} are preserved.

\begin{figure}
\includegraphics[width=.9 \columnwidth]{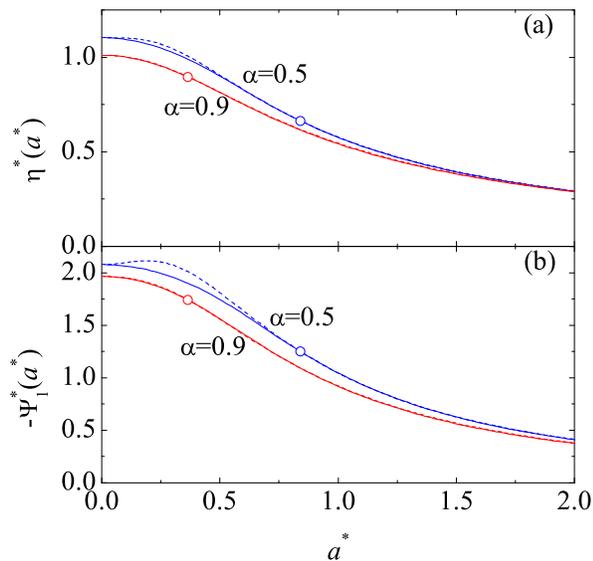}
\caption{(Color online) Shear-rate dependence of (a) the  viscosity function and (b) the first viscometric function for $\alpha=0.5$ and $\alpha=0.9$ in the USF with $d=3$. The solid lines have been obtained from numerical solutions of Eqs.\ \protect\eqref{3.12} and \protect\eqref{3.13}, while the dashed lines correspond to the simplified rheological model \protect\eqref{etaUSFPade} and \protect\eqref{Psi1}. The circles represent the steady-state values [cf.\ Eqs.\ \protect\eqref{3.9} and \protect\eqref{3.10}]. Note that the  simplified solution deviates practically from the numerical solution  only for the most inelastic system ($\alpha=0.5$) and in the region $a^*<a_s^*$.}
\label{fig4}
\end{figure}

\begin{figure}
\includegraphics[width=.9 \columnwidth]{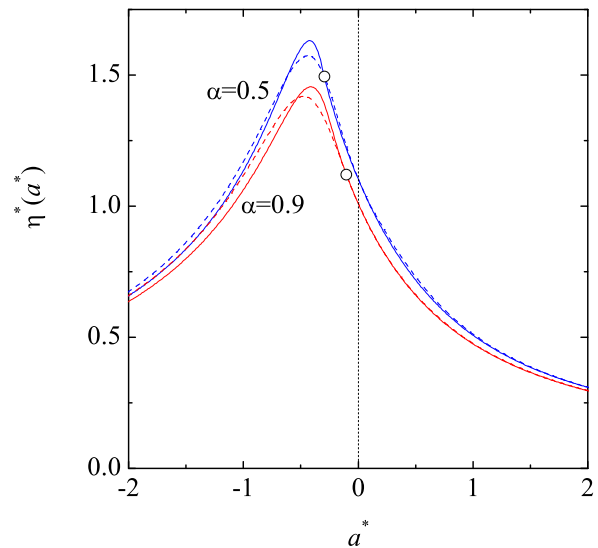}
\caption{(Color online) Longitudinal-rate dependence of the  viscosity  for $\alpha=0.5$ and $\alpha=0.9$ in the ULF with $d=3$. The solid lines have been obtained from numerical solutions of Eq.\ \protect\eqref{3.12}, while the dashed lines correspond to the simplified rheological model \protect\eqref{etaULF}. The circles represent the steady-state values [cf.\ Eqs.\ \protect\eqref{3.11a} and \protect\eqref{3.11b}].}
\label{fig5}
\end{figure}

Moreover, even in the physical case of hard spheres
($q=\frac{1}{2}$), Eqs.\ \eqref{etaULF}, \eqref{etaUSFPade}, and \eqref{Psi1} represent an excellent
analytical approximation to the numerical solution of the set of Eqs.\
\eqref{3.12} and \eqref{3.13} with appropriate initial conditions \cite{SGD04,S09}. The results for $\alpha=0.5$ and $\alpha=0.9$ are presented in Figs.\ \ref{fig4} and \ref{fig5} in the cases of the USF and the ULF, respectively.
We observe a generally good agreement between the numerical and the simplified results. This is especially true in the USF, where the limitations of the rheological model are only apparent for the most inelastic system ($\alpha=0.5$) and for shear rates smaller than the steady-state value, the agreement being better for the  viscosity than for the first viscometric function. In the ULF case the differences are more important, both for $\alpha=0.5$ and $\alpha=0.9$, although they are restricted to longitudinal rates near the maximum and also to values more negative than that of the maximum.

\section{Simulation details\label{sec4}}

We have performed DSMC simulations of the three-dimensional ($d=3$) USF and ULF. The details are similar to those described elsewhere \cite{AS05}, so here we provide only some distinctive features.

Since the original Boltzmann equation \eqref{2} is fully equivalent to the scaled form \eqref{19} [with the change of variables \eqref{11}--\eqref{15n}], we have solved the latter and have applied the  periodic boundary conditions \eqref{23}. We call this the ``inhomogeneous'' problem since the scaled distribution function $\of(\widetilde{\h},\ov,\ot)$ is, in principle, allowed to depend on the scaled spatial variable $\widetilde{\h}$. On the other hand, if one restricts oneself to \emph{uniform} solutions \eqref{22n}, then Eq.\ \eqref{19} reduces to Eq.\ \eqref{31}. The solution to Eq.\ \eqref{31} by the DSMC method will be referred to as the ``homogeneous'' problem. Most of the results that will be presented in Sec.\ \ref{sec5} correspond to the homogeneous problem.

A wide sample of initial conditions $\of_0(\widetilde{\h},\widetilde{\mathbf{v}})$ has been considered, as described below.
Note that, since $\dot{\theta}(0)=1$ [cf.\ Eq. \eqref{13}], $\widetilde{\h}=\h$ and $\widetilde{\mathbf{v}}=\mathbf{v}-a_0\h \mathbf{e}_x$ at $t=0$. Consequently, $\widetilde{n}_0(\widetilde{\h})=n_0(\h)$ and  $\widetilde{\mathbf{u}}_0(\widetilde{\h})=\mathbf{u}_0(\h)-a_0\h \mathbf{e}_x$. However, for consistency, we will keep the tildes in the expressions of the  initial state. An exception will be the initial temperature because $\widetilde{T}=T$ for all times.

The inhomogeneous problem for the USF was already analyzed in Ref.\ \cite{AS05} and, thus, only the inhomogeneous problem for the ULF is considered in this paper. The chosen initial condition is
\beq
\of_0(\widetilde{x},\widetilde{\mathbf{v}})=\frac{\on_0(\widetilde{x})}{4\pi}\frac{m}{3T_0}
\delta\left(|\widetilde{\mathbf{v}}-\widetilde{\mathbf{u}}_0(\widetilde{x})|-\sqrt{3T_0/m}\right),
\label{4.2}
\eeq
where  $\delta(x)$ is Dirac's distribution and the initial density and velocity fields are
\beq
\on_0(\widetilde{x})=\langle \on\rangle\left(1+\frac{1}{2}\sin\frac{2\pi \widetilde{x}}{\widetilde{L}}\right),
\label{4.3}
\eeq
\beq
\widetilde{\mathbf{u}}_0(\widetilde{x})=a_0\widetilde{L}\left(\cos\frac{\pi
\widetilde{x}}{\widetilde{L}}-\frac{2}{\pi}\right)\mathbf{e}_x,
\label{4.4}
\eeq
respectively, while the initial temperature $T_0$ is uniform.
In Eq.\ \eqref{4.3} $\langle \on\rangle$ is the  density spatially averaged between $\widetilde{x}=-\widetilde{L}/2$ and $\widetilde{x}=\widetilde{L}/2$. This quantity is independent of time.
In our simulations of the inhomogeneous problem we have taken $\alpha=0.5$ for the coefficient of restitution, $a_0=-4/\tau_0$  for the initial longitudinal rate, and $\widetilde{L}=2.5\lambda$ for the (scaled) box length. Here,
\beq
\lambda=\frac{1}{\sqrt{2}\pi\langle \on\rangle \sigma^2},\quad \tau_0=\frac{\lambda}{\sqrt{2T_0/m}}
\label{4.1}
\eeq
are a characteristic mean free path and an initial characteristic collision time, respectively.

\begin{table}
\caption{Values of the initial pressure tensor for the four initial conditions of the class \protect\eqref{4.6}.\label{table2}}
\begin{ruledtabular}
\begin{tabular}{cccccc}
Label&$\phi$&$\widetilde{P}_{xx,0}$&$\widetilde{P}_{yy,0}$&$\widetilde{P}_{zz,0}$&$\widetilde{P}_{xy,0}$\\
\hline
B0&$0$&$2\on T_0$&$0$&$\on T_0$&$0$\\
B1&$\pi/4$&$\on T_0$&$\on T_0$&$\on T_0$&$-\on T_0$\\
B2&$\pi/2$&$0$&$2\on T_0$&$\on T_0$&$0$\\
B3&$3\pi/4$&$\on T_0$&$\on T_0$&$\on T_0$&$\on T_0$\\
\end{tabular}
\end{ruledtabular}
\end{table}

As for the homogeneous problem (both for USF and UFL) two classes of initial conditions have been chosen. First, we have taken the local equilibrium state (initial condition A), namely
\beq
\of_0(\widetilde{\mathbf{v}})=\on\left(\frac{m}{2\pi T_0}\right)^{3/2}e^{-m\widetilde{v}^2/2T_0}.
\label{4.5}
\eeq
The other class of  initial conditions is of the anisotropic form
\begin{eqnarray}
\of_0(\widetilde{\mathbf{v}})&=&\frac{\on}{2}\left(\frac{m}{2\pi T_0}\right)^{1/2}
e^{-{m \widetilde{v}_{z}^{2}}/{2T_0}}\nn
&&\times\left[\delta\left(\widetilde{v}_{x}-V_{0}\cos
\phi\right)\delta\left(\widetilde{v}_{y}+V_{0}
\sin \phi\right)\right. \nonumber \\
&&\left.+\delta\left(\widetilde{v}_{x}+V_{0}\cos
\phi\right)\delta\left(\widetilde{v}_{y}-V_{0}\sin \phi\right)\right],
\label{4.6}
\end{eqnarray}
where $V_0\equiv\sqrt{2T_0/m}$
is the initial thermal speed and $\phi\in [0,\pi]$ is an  angle
characterizing each specific condition.   The pressure tensor
corresponding to Eq.\ \eqref{4.6} is given by
$\widetilde{P}_{xx,0}=2\on T_0\cos^2\phi$, $\widetilde{P}_{yy,0}=2\on T_0\sin^2\phi$,
$\widetilde{P}_{zz,0}=\on T_0$, and $\widetilde{P}_{xy,0}=-\on T_0\sin 2\phi$. The four values of
$\phi$ considered are $\phi=k \pi/4$ with $k=0$, $1$, $2$, and $3$;
we will denote the respective initial conditions of type \eqref{4.6}
as B0, B1, B2, and B3. The  values of the elements of the pressure tensor for these four initial conditions are displayed in Table \ref{table2}.
In the fully developed USF it is expected that $\widetilde{P}_{xy}<0$ (if $a>0$) and $\widetilde{P}_{xx}>\widetilde{P}_{yy}$. As we see from Table \ref{table2}, the four initial conditions are against those inequalities, especially in the case of  condition B3. As for the fully developed ULF, the physical expectations are $\widetilde{P}_{xy}=0$, $\widetilde{P}_{yy}=\widetilde{P}_{zz}$, and $\widetilde{P}_{xx}<\widetilde{P}_{yy}$ if $a_0>0$ and $\widetilde{P}_{xx}>\widetilde{P}_{yy}$ if $a_0<0$ [cf.\ Eq.\ \eqref{Pxx-Pyy}]. Again, none of the four initial conditions is consistent with those physical expectations, especially in the case of  condition B0 if $a_0>0$ and B2 if $a_0<0$.
The ``artificial'' character of the initial conditions \eqref{4.6} represents a stringent test of the scenario depicted in Fig.\ \ref{fig1}.

As summarized in Table \ref{table1}, when $\ozeta >2|a_0\oP_{xy}|/3\on\oT$ in the USF or when $\ozeta >-2a_0\oP_{xx}/3\on\oT$ in the ULF, the temperature
decreases with time (cooling states) either without lower bound (ULF with $a_0>0$) or until reaching the steady state (ULF with $a_0<0$ and USF). In those cooling states the temperature can decrease so much (relative to the initial value) that this might create technical problems (low signal-to-noise ratio), as mentioned at the end of Sec.\ \ref{sec2}. This can be corrected by the application of a \emph{thermostatting} mechanism, as represented by the change of variables \eqref{34} and \eqref{35} with  $\dot{\lc}(\ot)\propto \left[\widetilde{T}(\ot)\right]^{1/2}$. The DSMC implementation of Eqs.\ \eqref{38} and \eqref{mu} is quite simple. Let us denote by $\{\wv_i(\wt);i=1,\ldots,N\}$ the (rescaled) velocities  of the $N$ simulated particles at time $\wt$.
The corresponding rescaled temperature and shear (or longitudinal) rate are $\wT(\wt)$ and $\wa(\wt)$, respectively. During the time step $\delta\wt$
the velocities change  due to the action of the (deterministic) nonconservative external force $-m\wa(\wt) \widehat{v}_\h \mathbf{e}_x$ and
also  due to the (stochastic) binary collisions. Let us denote by $\{\wv_i'(\wt+\delta\wt);i=1,\ldots,N\}$ and by $\wT'(\wt+\delta\wt)$ the velocities and temperature after this stage. Thus, the action of the thermostat force $-m\mu(\wt)\wv$ is equivalent to the velocity rescaling
$\wv_i'(\wt+\delta\wt)\to \wv_i(\wt+\delta\wt)=\wv_i'(\wt+\delta\wt)\sqrt{\wT(\wt)/\wT'(\wt+\delta\wt)}$, so  $\wT'(\wt+\delta\wt)\to \wT(\wt+\delta\wt)=\wT(\wt)$. Similarly, the rescaled shear or longitudinal rate is updated as $\wa(\wt+\delta\wt)=\wa(\wt)\sqrt{\wT'(\wt+\delta\wt)/\wT(\wt)}$, so $\wa(\wt+\delta\wt)/\sqrt{\wT'(\wt+\delta\wt)}=\wa(\wt)/\sqrt{\wT(\wt)}$.

For each one of the five initial conditions for the homogeneous problem we have considered three coefficients of restitution: $\alpha=0.5$, $0.7$, and $0.9$. In the case of the USF, the values taken for the shear rate have been $a=0.01/\tau_0$, $a=0.1/\tau_0$, $a=4/\tau_0$, and $a=10/\tau_0$. The
two first values ($a=0.01/\tau_0$ and
$a=0.1/\tau_0$) are small enough to correspond to cooling cases,
even for the least inelastic system ($\alpha=0.9$), while the other
two values ($a=4/\tau_0$ and $a=10/\tau_0$) are large enough to
correspond to heating cases, even for the most inelastic system
($\alpha=0.5$). In the case of the ULF we have chosen $a_0=0.01/\tau_0$, $a_0=-0.01/\tau_0$, and $a_0=-10/\tau_0$. The first and second values correspond to cooling states (without and with a steady state, respectively), while the third value corresponds to  heating states.
Therefore, the total number of independent systems simulated in the homogeneous problem is 60 for the USF and 45 for the ULF.

The technical parameters of the simulations have been the following ones: $N=10^6$ simulated particles, an adaptive time step $\delta t=10^{-3}\tau_0\sqrt{T_0/\langle T\rangle}$, and a layer thickness (inhomogeneous problem) $\delta \widetilde{x}=0.05\lambda$.
Moreover, in order to improve the statistics, the results have been averaged over 100 independent realizations.

\section{Results\label{sec5}}

\subsection{USF. Homogeneous problem}

We have simulated the Boltzmann equation describing the homogeneous problem of the USF, i.e., Eq.\ \eqref{31} with $\h=y$, by means of the DSMC method. As said in Sec.\ \ref{sec4}, three coefficients of restitution ($\alpha=0.5$, $0.7$, and $0.9$) and four shear rates ($a=0.01/\tau_0$, $0.1/\tau_0$, $4/\tau_0$, and $10/\tau_0$) have been considered. For each combination of $\alpha$ and $a$, five different initial conditions (A and B0--B3) have been chosen.
In the course of the simulations we focus on the temporal evolution  of the  elements of the reduced pressure tensor $P_{ij}^*$, Eq.\ \eqref{3.6}, and of the reduced shear rate $a^*$, Eq.\ \eqref{3.5}. {}From these quantities one can evaluate the viscosity $\eta^*$, Eq.\ \eqref{etaUSF}, the first viscometric function $\Psi_1^*$, Eq.\ \eqref{43}, and the second viscometric function $\Psi_2^*$, Eq.\ \eqref{43bis}. The effective collision frequency  $\nu$ in Eq.\ \eqref{3.5} is defined by
\beq
\nu= {\frac{p}{\eta_{\text{NS}}^{\text{el}}}=}\frac{1}{1.016}\frac{16\sqrt{\pi}}{5}n\sigma^2\sqrt{\frac{T}{m}}.
\label{nuB}
\eeq
 {The factor $1.016$ comes from an elaborate Sonine approximation employed to determine the NS shear viscosity $\eta_{\text{NS}}^{\text{el}}$ of a gas of elastic hard spheres \cite{CC70}. For simplicity, and to be consistent with the approximate character of the kinetic model \eqref{n1}, this factor is not included in Eq.\ \eqref{0.1}.}

Note that the initial reduced shear rate is $a_0^*=a\tau_0/0.8885$.
Time is monitored through the accumulated number of collisions per particle, i.e., the total number of collisions in the system since the initial state, divided by the total number of particles. The reduced quantities and the number of collisions per particle are  not affected by the changes of variables discussed in Sec.\ \ref{sec2}.

\begin{figure}
\includegraphics[width=.9 \columnwidth]{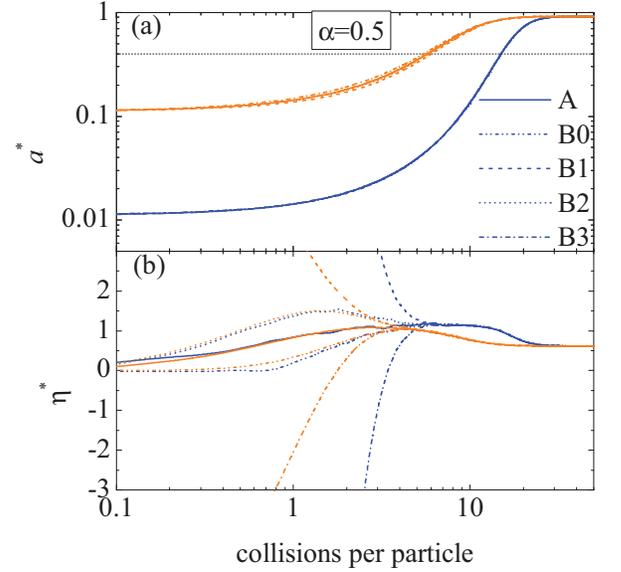}
\caption{(Color online) (a) Reduced shear rate $a^*$   and (b) reduced viscosity $\eta^*$  versus the number of collisions per particle  for the USF with $\alpha=0.5$ in the cooling states
 $a=0.01 /\tau_0$ [blue (dark gray) lines] and $a=0.1/\tau_0$ [orange (light gray) lines]. The legend refers to the five initial conditions considered. The dotted horizontal line in  panel (a) denotes the value $a_h^*=0.4$ above which the hydrodynamic regime is clearly established (see text).
\label{fig6}}
\end{figure}

\begin{figure}
\includegraphics[width=.9 \columnwidth]{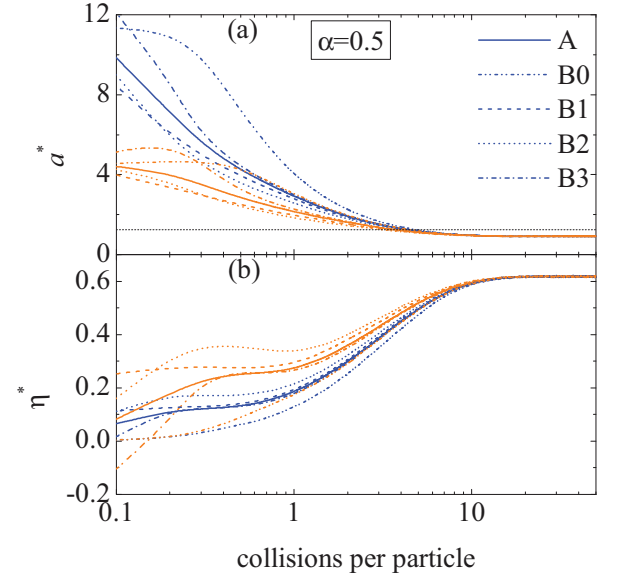}
\caption{(Color online) (a) Reduced shear rate $a^*$   and (b) reduced viscosity $\eta^*$ versus the number of collisions per particle  for the USF with $\alpha=0.5$ in the heating states
 $a=4 /\tau_0$ [orange (light gray) lines] and $a=10/\tau_0$ [blue (dark gray) lines]. The legend refers to the five initial conditions considered. The dotted horizontal line in  panel (a) denotes the value $a_h^*=1.25$ below which the hydrodynamic regime is clearly established (see text).
\label{fig7}}
\end{figure}

As a representative case, we first present results for the most inelastic system ($\alpha=0.5$).
Figures \ref{fig6} and \ref{fig7} show the evolution of $a^*$ and $\eta^*$ for the cooling states ($a=0.01/\tau_0$ and $0.1/\tau_0$) and the heating states ($a=4/\tau_0$ and $10/\tau_0$), respectively.
We  clearly observe that after about 30 collisions per particle (cooling states) or 20 collisions per particle (heating states) both $a^*$ and $\eta^*$ have  reached their stationary values.

Figure \ref{fig6} shows that, for each value of $a$, the full temporal evolution  of
$a^*\propto T^{-1/2}$ is practically independent of the
initial condition, especially in the case $a=0.01/\tau_0$. This is due to the fact that for these low values of $a\tau_0$ the
viscous heating term $-2aP_{xy}/dn$ in Eq.\ \eqref{5} can be neglected versus the inelastic cooling term $\zeta T$ for short times,
so the temperature initially evolves as in the homogeneous
cooling state (decaying practically exponentially with the number of
collisions), hardly affected by the details of the initial state. On the other hand, the first stage in the evolution of the reduced viscosity $\eta^*$ is widely dependent on the type of initial condition, as expected from the values of $\widetilde{P}_{xy,0}$ shown in Table \ref{table2}. In the heating cases, Fig.\ \ref{fig7} shows that the evolution followed by both $a^*$ and $\eta^*$ is distinct for each initial condition, except when the steady state is practically reached.

\begin{figure}
\includegraphics[width=.9 \columnwidth]{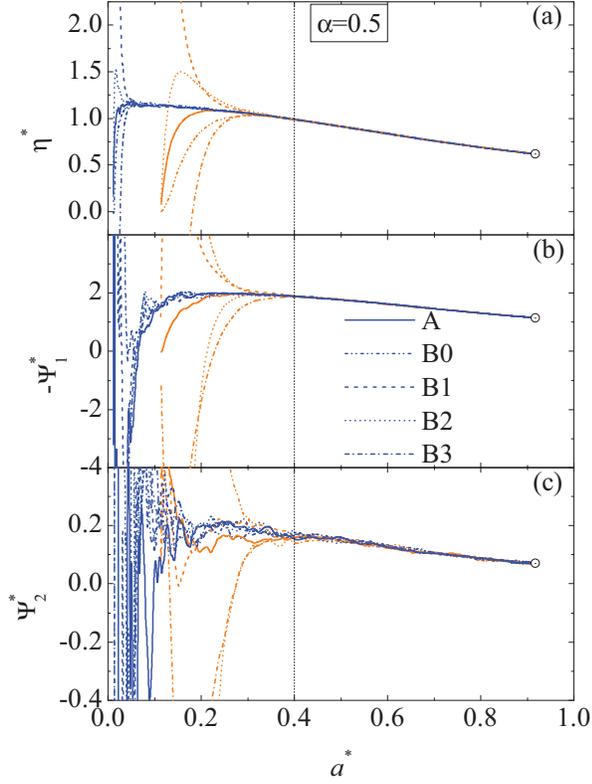}
\caption{(Color online) (a) Reduced viscosity $\eta^*$, (b) first viscometric function $-\Psi_1^*$, and (c) second viscometric function $\Psi_2^*$  versus the reduced shear rate $a^*$ for the USF with $\alpha=0.5$ in the cooling states
 $a=0.01 /\tau_0$ [blue (dark gray) lines] and $a=0.1/\tau_0$ [orange (light gray) lines]. The legend refers to the five initial conditions considered. The circles represent the steady-state points $(a_s^*,\eta_s^*)$, $(a_s^*,-\Psi_{1,s}^*)$, and $(a_s^*,\Psi_{2,s}^*)$, respectively. The dotted vertical line denotes the value $a_h^*=0.4$ above which the curves collapse to a common one.
\label{fig8}}
\end{figure}

\begin{figure}
\includegraphics[width=.9 \columnwidth]{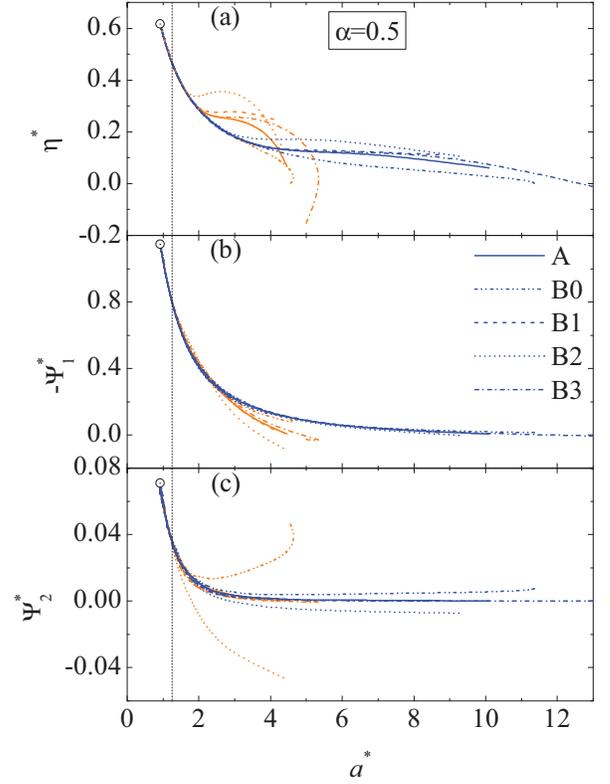}
\caption{(Color online) (a) Reduced viscosity $\eta^*$, (b) first viscometric function $-\Psi_1^*$, and (c) second viscometric function $\Psi_2^*$ versus the reduced shear rate $a^*$ for the USF with $\alpha=0.5$ in the heating states
 $a=10 /\tau_0$ [blue (dark gray) lines] and $a=4/\tau_0$ [orange (light gray) lines]. The legend refers to the five initial conditions considered. The circles represent the steady-state points $(a_s^*,\eta_s^*)$, $(a_s^*,-\Psi_{1,s}^*)$, and $(a_s^*,\Psi_{2,s}^*)$, respectively. The dotted vertical line denotes the value $a_h^*=1.25$ below which the curves collapse to a common one.
\label{fig9}}
\end{figure}

In any case, the interesting point is whether an unsteady hydrodynamic regime is established \emph{prior} to the steady state.
If so,  a parametric plot of $\eta^*$ versus $a^*$ must approach a well-defined function $\eta^*(a^*)$, regardless of the initial
condition. Figures \ref{fig8} and \ref{fig9} present such a parametric plot, also for the viscometric functions, for the cooling and heating states, respectively.
We observe that, for each class of states (either cooling or heating) , the 10 curves are \emph{attracted} to a common smooth ``universal'' curve, once the kinetic stage (characterized by strong variations, especially in the case of the second viscometric function for the cooling states) is over. One can safely say that the hydrodynamic regime extends to the range $0.4\lesssim a^*\leq a^*_s$ for the considered cooling states  and to the range $1.25\gtrsim a^*\geq a_s^*$ for the considered heating states. We will denote the above threshold values of $a^*$ by $a_h^*$. It is expected that $a_h^*$ depends on the initial value $a_0^*$ (apart from a weaker dependence on the details of the initial distribution); in fact, Figs.\ \ref{fig8} and \ref{fig9} show that the hydrodynamic regime is reached at a value $a_h^*<0.4$ by the states with $a=0.01/\tau_0$ and at a value $a_h^*>1.25$ by the states with $a=10/\tau_0$. Here, however, we adopt a rather conservative criterion and take a common value $a_h^*=0.4$ for $a=0.01/\tau_0$ and $0.1\tau_0$ and a common value $a_h^*=1.25$ for $a=4/\tau_0$ and $10\tau_0$. It is also quite apparent from Figs.\ \ref{fig8} and \ref{fig9} that the collapse to a common curve takes place earlier for $\eta^*$ than for $\Psi_1^*$, $\Psi_2^*$ being the quantity with the largest ``aging'' period.

\begin{table}
\caption{This table shows, for each value of the coefficient of restitution $\alpha$ and each value of $a\tau_0$, the duration of the aging period toward the unsteady hydrodynamic regime and the total duration of the transient period toward the steady state, both measured by the number of collisions per particle.    The aging period is defined as the number of collisions per particle needed to reach a common threshold value $a_h^*$ above (below) which the hydrodynamic regime is established for the cooling (heating) states. The table also includes the stationary value $a_s^*$ of the reduced shear rate.\label{table3}}
\begin{ruledtabular}
\begin{tabular}{cccccc}
$\alpha$&$a_s^*$&$a\tau_0$&$a_h^*$&Aging period&Transient period\\
\hline
$0.5$&$0.92$&$0.01$&$0.4$&$15$&$30$\\
&&$0.1$&$0.4$&$5$&$20$\\
&&$4$&$1.25$&$5$&$20$\\
&&$10$&$1.25$&$5$&$20$\\
\hline
$0.7$&$0.68$&$0.01$&$0.35$&$20$&$50$\\
&&$0.1$&$0.35$&$8$&$30$\\
&&$4$&$1$&$5$&$30$\\
&&$10$&$1$&$5$&$30$\\
\hline
$0.9$&$0.37$&$0.01$&$0.2$&$50$&$100$\\
&&$0.1$&$0.2$&$10$&$60$\\
&&$4$&$0.8$&$5$&$50$\\
&&$10$&$0.8$&$5$&$50$\\
\end{tabular}
\end{ruledtabular}
\end{table}

{}From Fig.\ \ref{fig6} it can be seen that the value $a_h^*=0.4$ is reached after about $5$ collisions per particle in the states with $a=0.1/\tau_0$ and after about 15 collisions per particle in the states with $a=0.01/\tau_0$. Similarly, Fig.\ \ref{fig7} shows that the value $a_h^*=1.25$ is reached after about $5$ collisions per particle in the states with $a=4/\tau_0$ and $a=10/\tau_0$. Given that, as said before, the values $a_h^*=0.4$ and $a_h^*=1.25$ are conservative estimates, we find that, as expected, the duration of the kinetic stage is  shorter than the duration of the subsequent hydrodynamic stage, before the steady state is eventually attained.

\begin{figure}
\includegraphics[width=.9 \columnwidth]{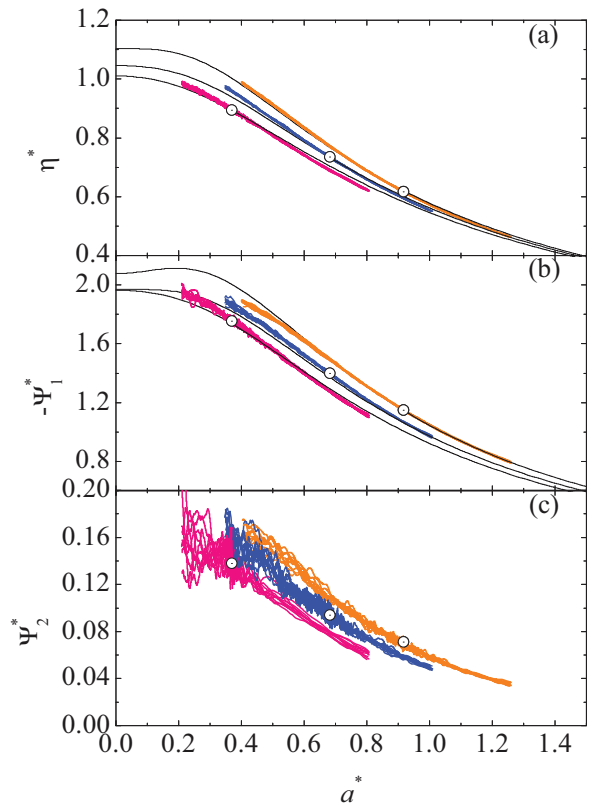}
\caption{(Color online) (a) Reduced viscosity $\eta^*$, (b) first viscometric function $-\Psi_1^*$, and (c) second viscometric function $\Psi_2^*$ versus the reduced shear rate $a^*$ for the USF with, from top to bottom, $\alpha=0.5$, $\alpha=0.7$, and $\alpha=0.9$. In order to focus on the hydrodynamic regime, the curves have been truncated at the values of $a_h^*$ given by Table \protect\ref{table3}. The circles represent the steady-state points. The thin solid lines in panels (a) and (b) represent the predictions of our simplified rheological model, Eqs.\ \protect\eqref{etaUSFPade} and  \protect\eqref{Psi1}. Note that the model is unable to predict a non-zero second viscometric function.
\label{fig10}}
\end{figure}

We have observed  behaviors similar to those of Figs.\ \ref{fig6}--\ref{fig9}
for the other two coefficients of restitution  ($\alpha=0.7$ and $\alpha=0.9$, not shown). Table \ref{table3} gives the values of $a_h^*$ and the duration of the aging and transient periods for the 12 classes of states analyzed. It turns out that the number of
collisions per particle the system needs to lose memory of its initial state is
practically independent of $\alpha$ for the heating states. However, the total duration of the
transient period increases with $\alpha$ \cite{AS07}. In fact, there is no true steady state in the elastic limit $\alpha\to 1$.   Therefore, the less inelastic the system, the smaller the fraction of the transient period (as measured by the number of collisions per particle) spent by the heating states in the kinetic regime. In the
cooling cases, both the aging and the transient periods increase with $\alpha$.

Figure \ref{fig10} displays the viscosity $\eta^*(a^*)$ and the viscometric functions $-\Psi_1^*(a^*)$ and $\Psi_2^*(a^*)$ for $\alpha=0.5$, $0.7$, and $0.9$, both for the cooling and the heating states. Here we have focused on the ranges of $a^*$ where the hydrodynamic regime can safely be assumed to hold, namely $0.4\leq a^*\leq 1.25$ for $\alpha=0.5$, $0.35\leq a^*\leq 1$ for $\alpha=0.7$, and $0.2\leq a^*\leq 0.8$ for $\alpha=0.9$. The curves describing the predictions of the simplified rheological model for $\eta^*$, Eq.\ \eqref{etaUSFPade}, and  for $\Psi_1^*$, Eq.\ \eqref{Psi1}, are also included. We can see that the curves corresponding to the cooling states ($a^*<a^*_s$) and those corresponding to the heating states ($a^*>  a^*_s$) smoothly match at the steady-state point. In the case of the nonlinear viscosity function $\eta^*$, the 10 curves building each branch (cooling or heating) for each value of $\alpha$  exhibit a very high degree of overlapping. Due to fluctuations associated with the normal stress differences, the common hydrodynamic curves for the viscometric functions are much more coarse grained, especially in the case of $\Psi_2^*$, whose magnitude is at least 10 times smaller than that of $\Psi_1^*$. The impact of fluctuations is higher in the cooling branches ($a^*<a_s^*$) than in the heating branches ($a^*>a_s^*$). In fact, the definition of the viscometric functions [see Eqs.\ \eqref{43} and \eqref{43bis}] shows that the signal-to-noise ratio is expected to deteriorate as the reduced shear rate $a^*$ decreases. Since $\eta^*$, $-\Psi_1^*$, and $\Psi_2^*$ decrease with decreasing inelasticity, the role played by fluctuations increase as $\alpha$ increases.
It is also interesting to remark that, despite its simplicity and analytical character,  the rheological model described by Eqs.\ \eqref{etaUSFPade} and  \eqref{Psi1} describes very well the nonlinear dependence of $\eta^*(a^*)$ and $\Psi_1^*(a)^*$. On the other hand, the simple kinetic model \eqref{n1} does not capture any difference between the normal stresses $P_{yy}$ and $P_{zz}$ in the hydrodynamic regime and, thus, it predicts a vanishing second viscometric function [cf.\ Eq.\ \eqref{Psi2_0}].

Figure \ref{fig10} strongly supports Eq.\ \eqref{10} in the USF, {i.e.},
the existence of well-defined hydrodynamic rheological functions
$P_{ij}^*(a^*)$ [or, equivalently, $\eta^*(a^*)$ and $\Psi_{1,2}^*(a^*)$] acting as ``attractors''  in the evolution of the
pressure tensor $P_{ij}(t|f_0)$, regardless of the initial
preparation $f_0$. The stronger statement \eqref{7} (see Fig.\ \ref{fig1}) is also supported by the simulation results \cite{AS07}.

\subsection{ULF. Inhomogeneous problem}

\begin{figure}
\includegraphics[width=.9 \columnwidth]{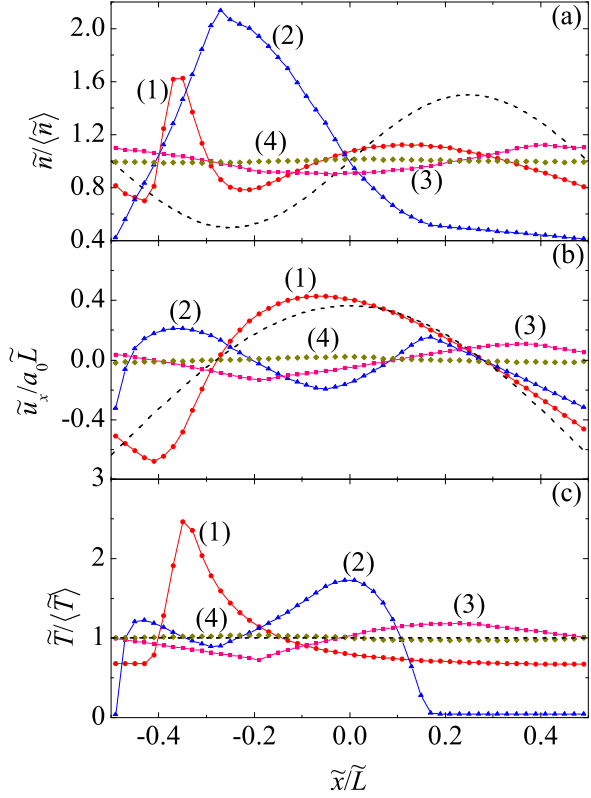}
\caption{(Color online) Profiles of (a) density, (b) flow velocity, and (c) temperature in the ULF, starting from the initial condition of Eqs.\ \protect\eqref{4.2}--\protect\eqref{4.4}. The curves correspond to (1) $\ot=0.046 \tau_0$ ($\approx 0.03~\text{coll}/\text{part}$), (2) $\ot=0.158 \tau_0$ ($\approx 0.29~\text{coll}/\text{part}$), (3) $\ot=0.306 \tau_0$ ($\approx 0.72~\text{coll}/\text{part}$), and (4) $\ot=0.400 \tau_0$ ($\approx 1.07~\text{coll}/\text{part}$). The dashed lines represent the initial profiles.
  \label{fig11}}
\end{figure}

Now we turn to the ULF sketched by Fig.\ \ref{fig3}. By performing the changes of variables \eqref{11}, \eqref{14}, and \eqref{15n} (with $\h=x$), we have numerically solved the Boltzmann equation \eqref{19} by means of the DSMC method. Although Eq.\ \eqref{19} admits homogeneous solutions in the fully developed ULF [see Eq.\ \eqref{22n}], it is worth checking that Eq.\ \eqref{19}, complemented by the periodic boundary conditions \eqref{23}, indeed leads an inhomogeneous initial state toward a (time-dependent) homogeneous state. A similar test was carried out in the case of the USF in Ref.\ \cite{AS05}.

As described in Sec.\ \ref{sec4}, we have considered the highly inhomogeneous initial state given by Eqs.\ \eqref{4.2}--\eqref{4.4} with $a_0=-4/\tau_0$ and $\widetilde{L}=2.5\lambda$, and solved Eq.\ \eqref{19} for a coefficient of restitution $\alpha=0.5$. The instantaneous density, flow velocity, and temperature profiles are plotted in Fig.\ \ref{fig11} at four representative times.
In order to decouple the relaxation to a homogeneous state from the increase of the global temperature (here viscous heating prevails over inelastic cooling), panel (c) of Fig.\ \ref{fig11} displays the ratio $\widetilde{T}/\langle \widetilde{T}\rangle$, with $\langle \widetilde{T}\rangle=\langle \widetilde{p}\rangle/\langle \widetilde{n}\rangle$, where $\langle \widetilde{n}\rangle$ and $\langle \widetilde{p}\rangle$ are the density and hydrostatic pressure, respectively, averaged across the system.

We observe that at $\ot=0.046 \tau_0$ the density, velocity, and temperature profiles are still reminiscent of the initial fields, except in the region $-0.4\lesssim \widetilde{x}/\widetilde{L}\lesssim -0.3$, where the density and the temperature present a maximum and the flow velocity exhibits a local minimum. At $\ot=0.158 \tau_0$ the inhomogeneities are still quite strong, with a widely depopulated region $0.2\lesssim \widetilde{x}/\widetilde{L}\leq 0.5$ of particles practically moving with the local mean velocity (which implies an almost zero temperature). By  $\ot=0.306 \tau_0$ the profiles have smoothed out significantly. Finally, the system becomes practically homogeneous at  $\ot=0.400 \tau_0$. Thus, the relaxation to the homogeneous state lasts about $1$ collision per particle only. The system keeps evolving to the steady state,  {which} requires about $20$ collisions per particle. In fact, $\langle \widetilde{T}\rangle\simeq 58 T_0$ after $1$ collision per particle, while $\langle\widetilde{T}\rangle\simeq 211 T_0$ in the steady state. Recall that $\widetilde{\mathbf{u}}(\widetilde{x},\ot)=0$ translates into $\mathbf{u}(x,t)=a(t)x\mathbf{e}_x$ [see Eq.\ \eqref{21n}].

It is important to remark that the relaxation to the ULF geometry observed in Fig.\ \ref{fig11} does not discard the possibility of spatial instabilities for sufficiently large values of the system size $\widetilde{L}$, analogously to what happens in the USF case \cite{S92,SK94,GT96,K00a,K00b,K01,G06}.

\subsection{ULF. Homogeneous problem}

\begin{figure}
\includegraphics[width=.9 \columnwidth]{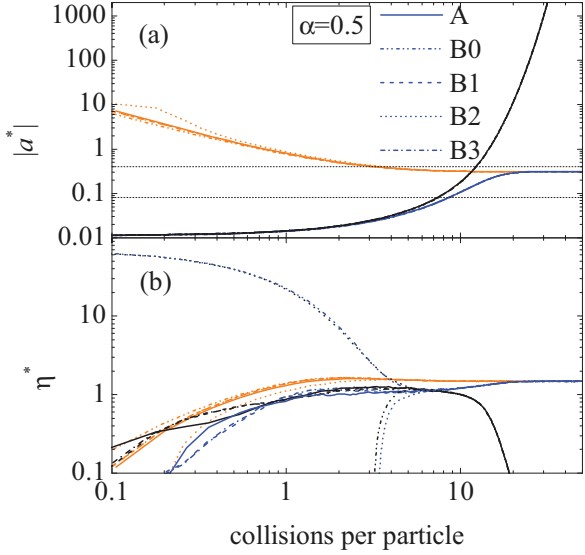}
\caption{(Color online) (a) Absolute value of  the reduced longitudinal rate $|a^*|$   and (b)  reduced viscosity $\eta^*$ versus the number of collisions per particle  for the ULF with $\alpha=0.5$ in the cooling states
$a_0=-0.01 /\tau_0$ [blue (dark gray) lines] and $a_0=0.01/\tau_0$ (black lines) and in the heating states $a_0=-10/\tau_0$ [orange (light gray) lines]. The legend refers to the five initial conditions considered. The dotted horizontal lines in  panel (a) denote the values $a_h^*=\pm 0.08$ and $a_h^*=-0.4$ (see text).
\label{fig12}}
\end{figure}

\begin{figure}
\includegraphics[width=.9 \columnwidth]{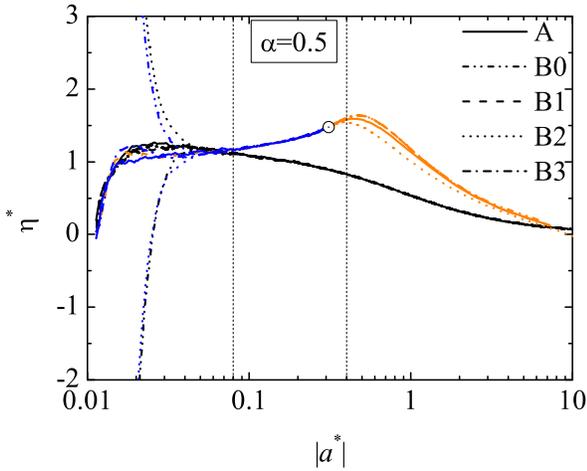}
\caption{(Color online) Reduced viscosity $\eta^*$  versus the absolute value of the reduced longitudinal rate $|a^*|$ for the ULF with $\alpha=0.5$ in the cooling states
$a_0=-0.01 /\tau_0$ [blue (dark gray) lines] and $a_0=0.01/\tau_0$ (black lines) and in the heating states $a_0=-10/\tau_0$ [orange (dark gray) lines]. The legend refers to the five initial conditions considered. The circle represents the steady-state point $(|a_s^*|,\eta_s^*)$. The dotted vertical lines  denote the values $a_h^*=\pm 0.08$ and $a_h^*=-0.4$ (see text).
\label{fig13}}
\end{figure}

Now we restrict to the homogeneous ULF problem. The homogeneous Boltzmann equation \eqref{31} wit $\h=x$ has been solved via the DSMC method outlined in Sec.\ \ref{sec4} for $\alpha=0.5$, $0.7$, and $0.9$, starting from the initial conditions \eqref{4.5} and \eqref{4.6} with $a_0=-10/\tau_0$, $-0.01/\tau_0$, and $0.01/\tau_0$. Note that the B1 initial state becomes  B3, and vice versa, under the change $v_y\to -v_y$, and so both are formally equivalent in the ULF geometry. As said before, a steady state is only possible if $a_0<0$. Moreover, the choices $a_0=\pm 0.01/\tau_0$ correspond to cooling states, while the choice $a_0=-10/\tau_0$ corresponds to heating states. In the course of the simulations the reduced longitudinal rate $a^*$, Eq.\ \eqref{3.5}, and the reduced nonlinear viscosity $\eta^*$, Eq.\ \eqref{Pxx-Pyy}, are evaluated.

As in the USF case, let us adopt $\alpha=0.5$ to illustrate the behaviors observed. Figure \ref{fig12} displays the time evolution of $|a^*|$ and $\eta^*$ for the 15 simulated states (5 for each value of $a_0$). The states with negative $a_0$ become stationary after about $20$ collisions per particle, a value comparable to what we observed in the USF case for $\alpha=0.5$ (see Table \ref{table3}). As for the states with $a_0=0.01/\tau_0$, $a^*$ monotonically increases and $\eta^*$ monotonically decreases (except for a possible transient stage) without bound. It is worth noticing that the first stage of evolution (up to about $7$ collisions per particle) of $|a^*|$ and $\eta^*$ for the initial condition B0 (B2) with $a_0=-0.01/\tau_0$  is very similar to those for the initial condition B2 (B0) with $a_0=0.01/\tau_0$.

Eliminating time between $a^*$ and $\eta^*$ one obtains the parametric plot shown in Fig.\ \ref{fig13}. In the cooling states $a_0=\pm 0.01/\tau_0$ we observe that the hydrodynamic regime is reached at approximately $|a_h^*|=0.08$. {}From Fig.\ \ref{fig12} we see that this corresponds to  about seven to eight collisions per particle. The heating states $a_0=-10/\tau_0$ deserve some extra comments. In those cases the time evolution is so rapid that, strictly speaking, the collapse of the five curves takes place only for $a_s^*\geq a^*\geq a_h^*=-0.4$,  {which} corresponds to an aging period of about three to four collisions per particle (see Fig.\ \ref{fig12}). On the other hand, it can be clearly seen from Fig.\ \ref{fig13} that the three curves corresponding to the initial states B0, B1, and B3 have collapsed much earlier and are not distinguishable on the scale of the figure. It seems that the isotropic local equilibrium initial state A  and the highly artificial anisotropic initial state B2  require a longer kinetic stage than in the cases of the initial states B0, B1, and B3.
{While the relatively slower convergence of the initial condition B2 can be expected because of its associated  ``incorrect'' negative viscosity, it seems paradoxical that the local equilibrium initial condition A also relaxes more slowly than the initial conditions B1 and B3, the three of them having a zero initial viscosity. This might be due to the isotropic character of the local equilibrium distribution, in contrast to the high anisotropy of conditions B1 and B3.}
In what follows we will discard the initial conditions A and B2 for $a_0=-10/\tau_0$ and assume that the states starting from the initial conditions B0, B1, and B3 have already reached the hydrodynamic stage for say $a^*\geq -2$. A stricter limitation to $a^*>-0.4$ would miss the interesting maximum of $\eta^*$ at $a^*<a_s^*$ predicted by the BGK-like kinetic model (see Fig.\ \ref{fig5}). In any case, as observed in Fig.\ \ref{fig13}, the behavior of the curves with $a_0=-10/\tau_0$ corresponding to the initial conditions B2 and, especially, A is very close to the one corresponding to the initial conditions B0, B1, and B3.

\begin{figure}
\includegraphics[width=.9 \columnwidth]{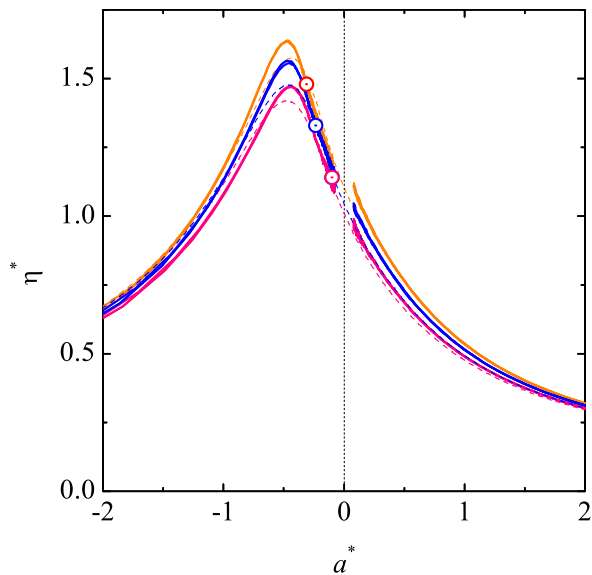}
\caption{(Color online) Reduced viscosity $\eta^*$   versus the reduced longitudinal rate $a^*$ for the ULF with, from top to bottom, $\alpha=0.5$, $\alpha=0.7$, and $\alpha=0.9$.  The circles represent the steady-state points. The thin dashed lines  represent the predictions of our simplified rheological model, Eq.\ \protect\eqref{etaULF}.
\label{fig14}}
\end{figure}

The analysis for the cases $\alpha=0.7$ and $0.9$ is similar to the one for $\alpha=0.5$ and, thus, it is omitted here. As in the USF (see Table \ref{table3}), the main effect of increasing $\alpha$ is to slow down the dynamics: the steady state (if $a_0<0$) is reached after a larger number of collisions and  the hydrodynamic stage requires a longer period.

The $a^*$-dependence of $\eta^*$ for the three values of $\alpha$ is shown in Fig.\ \ref{fig14}, where we have focused on the intervals $-2\leq a^*\leq a_s^*$ for $a_0=-10/\tau_0$ (initial conditions B0, B1, and B3), $-0.08\geq a^*\geq a_s^*$ for $a_0=-0.01/\tau_0$ (initial conditions A and B0--B3), and $ a^*\geq 0.08$ for $a_0=0.01/\tau_0$ (initial conditions A and B0--B3). Analogously to the case of Fig.\ \ref{fig10}, it can be seen that the heating and cooling branches with negative $a_0$ smoothly match at the steady state. Moreover, the cooling branch with positive $a_0$ is a natural continuation of the cooling branch with negative $a_0$, even though the zero longitudinal rate $a^*=0$ represents a repeller in the time evolution of both branches. Figure \ref{fig14} also includes the predictions of the rheological model \eqref{etaULF}. The agreement with the simulation results is quite satisfactory, although the model tends to underestimate the maxima. This discrepancy is largely corrected by the true numerical solution of Eq.\ \eqref{3.12} (see Fig.\ \ref{fig5}). However, as done in Fig.\ \ref{fig10}, we prefer to keep the rheological model due to its explicit and analytical character.

\section{Conclusions\label{sec6}}

In this paper we have investigated whether  the scenario of \emph{aging to
hydrodynamics} depicted in Fig.\ \ref{fig1} for conventional gases applies to granular gases as well, even at
high dissipation. Here the term \emph{hydrodynamics} means that the velocity distribution function, and, hence, the irreversible fluxes, is a functional of the hydrodynamic fields (density, flow velocity, and granular temperature) and, thus, it is not limited to the NS regime. To address the problem, we have restricted ourselves to unsteady states in two classes of flows, the USF and the ULF (see Figs.\ \ref{fig2} and \ref{fig3}). While the USF is an incompressible flow ($\nabla\cdot\mathbf{u}=0$) and the ULF is a compressible one ($\nabla\cdot\mathbf{u}\neq 0$), they share some physical features (uniform density, temperature, and rate of strain tensor) that allow for a unified theoretical framework.  Both flows admit heating states (where viscous heating prevails over inelastic cooling) and cooling states (where inelastic cooling overcomes viscous heating), until a steady state is eventually reached. Moreover, in the ULF with positive longitudinal rates only ``super-cooling'' states (where inelastic cooling adds to ``viscous cooling'') are possible and, thus, no steady state exists.

Two complementary routes have been adopted. First, the BGK-like model kinetic equation \eqref{n1} has been used in lieu of the true inelastic Boltzmann equation, which allows one to derive a closed set of nonlinear first-order differential equations [cf.\ Eq.\ \eqref{3.2bis}] for the temporal evolution of the elements of the pressure tensor. A numerical solution with appropriate initial conditions and elimination of time between the reduced pressure tensor $P_{ij}^*$ and the reduced rate of strain $a^*$ provides the \emph{non-Newtonian} functions $P_{ij}^*(a^*)$, from which one can construct the viscosity function $\eta^*(a^*)$ [cf.\ Eq.\ \eqref{eta}] and (only in the USF case) the viscometric functions  $\Psi_1^*(a^*)$ [cf.\ Eq.\ \eqref{43}] and $\Psi_2^*(a^*)$ [cf.\ Eq.\ \eqref{43bis}]. The numerical task can be avoided at the cost of introducing approximations. The one we have worked out consists of expanding the solution in powers of a parameter $q$ measuring the hardness of the interaction ($q=0$ for inelastic Maxwell particles and $q=\frac{1}{2}$ for inelastic hard spheres), truncating the expansion to first order, and then constructing Pad\'e approximants. This yields explicit expressions for the rate of strain dependence of the rheological functions. In the USF case the viscosity and the first viscometric functions are given by Eqs.\ \eqref{etaUSFPade} and \eqref{Psi1}, respectively, complemented by Eqs.\ \eqref{15}, \eqref{hUSF}, and \eqref{HUSF}; the second viscometric function vanishes in the BGK-like kinetic model \eqref{n1}. As for the ULF, only one rheological function (viscosity) exists and it is given by Eq.\ \eqref{etaULF}, complemented by Eqs.\ \eqref{15ULF} and \eqref{hULF}. While one could improve the approximation by including terms in the $q$ expansion beyond the first-order one \cite{S00a},  the approximation considered in this paper represents a balanced compromise between simplicity and accuracy (see Figs.\ \ref{fig4} and \ref{fig5}). In fact, the results predicted by the kinetic model \eqref{n1} and the simplified rheological model described by Eqs.\ \eqref{etaULF}, \eqref{etaUSFPade}, and \eqref{Psi1} share the non-Newtonian solution for $q=0$ as well as the  steady-state values and the values at $a^*=0$ for arbitrary $q$.

The second, and most important, route has been the numerical solution of the true Boltzmann equation by the stochastic DSMC method for three values of the coefficient of restitution and a wide sample of initial conditions. The most relevant results are summarized in Figs.\ \ref{fig10} (USF) and \ref{fig14} (ULF). Those figures show an excellent degree of overlapping of the rheological functions obtained by starting from the different initial conditions, although the USF viscometric functions may exhibit large fluctuations. The overlapping takes place after a kinetic stage lasting about $5$ collisions per particle for the heating states considered and between $5$ and $50$ collisions per particle for the cooling states considered (see Table \ref{table3}). In the latter states the whole dynamics is slowed down with respect to the heating states, so  the increase of the duration of the kinetic stage is correlated with a similar increase of the duration of the subsequent hydrodynamic stage. We have also observed that the characteristic time periods  increase as the inelasticity decreases. Figures \ref{fig10} and \ref{fig14} also show that the rheological model, despite its simplicity, captures reasonably well, even at a quantitative level, the main features of the DSMC results. An exception is the USF second viscometric function which, although about 10 times smaller than the first viscometric function, is unambiguously nonzero.

In summary, we believe that our results provide strong extra support to the validity of a hydrodynamic description of granular gases outside the quasielastic limit and the NS regime.
 {Of course, it is important to bear in mind that the USF and ULF  are special classes of flows where no density or thermal gradients exist, except during the early kinetic stage (see, for instance, Fig.\ \ref{fig11}). Thus, the results presented here do not  guarantee \emph{a priori} the applicability of a non-Newtonian hydrodynamic approach for a general situation in the presence of density and thermal gradients. On the other hand, recent investigations for Couette--Fourier flows \cite{BKR09,VSG10,VGS11,VGS11b,TTMGSD01} nicely complement the study presented in this paper in favor of a (non-Newtonian) hydrodynamic treatment of granular gases.}

\acknowledgments
The authors  acknowledge support from the Ministerio de Ciencia e Innovaci\'on (Spain) through Grant No.\ FIS2010-16587 and from the Junta de Extremadura (Spain) through Grant No.\ GR10158, partially financed by FEDER (Fondo Europeo de Desarrollo Regional) funds.

\ed